# Multimodal Multi-Agent Ransomware Analysis Using AutoGen

Aimen Wadood [1,2], Mubashar Iqbal [1,2], Umme Zahoora [4], Asifullah Khan [1,2,3]

[1] Pattern Recognition Lab, DCIS, PIEAS, Nilore, Islamabad, Pakistan.

[2] Deep Learning Lab, Center for Mathematical Sciences, PIEAS, Nilore, Islamabad, Pakistan.

[3] PIEAS Artificial Intelligence Center (PAIC), PIEAS, Nilore, Islamabad, Pakistan.

[4] Department of Creative Technology, AIR University, E9 Campus, Islamabad, Pakistan.

*Corresponding author(s). E-mail(s): asif@pieas.edu.pk

## Abstract

Ransomware has become one of the most serious cybersecurity threats causing major financial losses and operational disruptions worldwide. Traditional detection methods such as static analysis, heuristic scanning and behavioral analysis often fail when used alone. To address these limitations, this paper presents multimodal multi agent ransomware analysis framework designed for ransomware classification.

Proposed multimodal multiagent architecture combines information from static, dynamic and network sources. Each data type is handled by specialized agent that uses auto encoder based feature extraction. These representations are then integrated through a fusion agent. After that fused representation are used by transformer based classifier. It identifies the specific ransomware family. The agents interact through an interagent feedback mechanism that iteratively refines feature representations by suppressing low confidence information. The framework was evaluated on large scale datasets containing thousands of ransomware and benign samples. Multiple experiments were conducted on ransomware dataset. It outperforms single modality and non-adaptive fusion baseline achieving improvement of up to 0.98 in Macro-F1 for family classification and reducing calibration error.

Over 100 epochs, the agentic feedback loop displays a stable monotonic convergence leading to over +0.75 absolute improvement in terms of agent quality and a final composite score of around 0.88 without fine tuning of the language models. Zeroday ransomware detection remains family dependent on polymorphism and modality disruptions. Confidence aware abstention enables reliable real world deployment by favoring conservative and trustworthy decisions over forced classification. The findings indicate that proposed approach provides a practical and effective path toward improving real world ransomware defense systems.

# 1. Introduction

Ransomware has become one of the most crucial cybersecurity threats causing billions of dollars in losses annually. It continues to evolve in frequency, diversity and sophistication across families such as Dharma, WannaCry, Locky and Revil etc. Severity based categorizations of ransomware [1][2] include scareware and fake alerts that threaten victims to pay for purportedly fake repair software [3]. Locker ransomware locks out the operating system and uses threatening rhetoric related to law enforcement [4]. Crypto ransomware encrypts user files using strong or hybrid cryptographic algorithms making recovery without payment difficult. Modus operandi of ransomware is depicted in Figure 1.

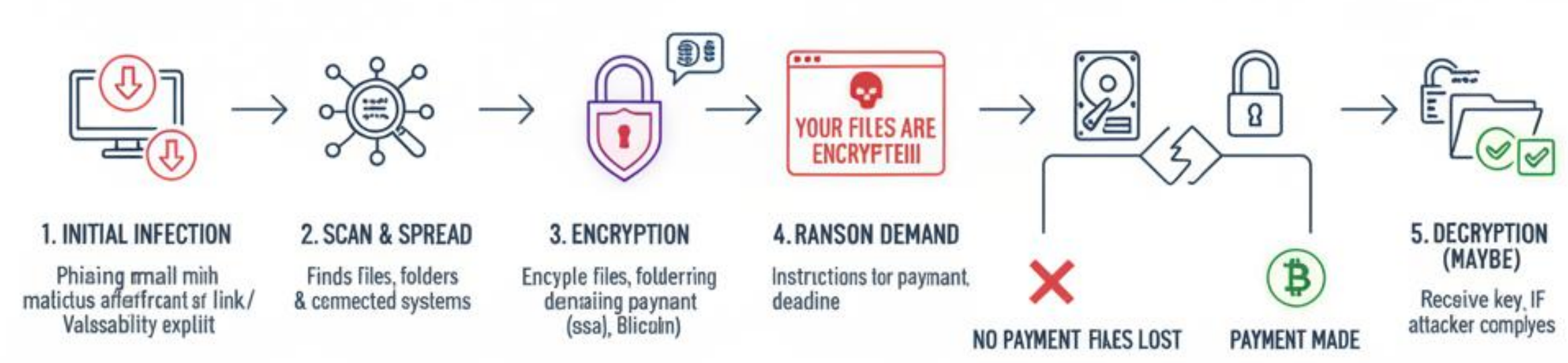

**Figure 1: Ransomware Plan of Action**

Conventional techniques for the detection of ransomware include signature based detection, static analysis, dynamic analysis, behavior based detection, heuristic based detection, access pattern monitoring, network traffic analysis and integrity checking etc. Signature based technique identifies ransomware by matching files against known ransomware signatures. Static analysis involves the study of program attributes without the need for execution which includes analysis of file headers, opcodes and import details, while dynamic analysis involves the execution of code inside sandbox environments [43]. Behavior based detection technique monitors suspicious actions such as mass file encryption or renaming the files etc. Heuristic methods use predefined rules and thresholds to flag ransomware activities. Access pattern monitoring detects abnormal file access and modification patterns by the ransomware. Lastly, network analysis and monitoring focuses on the observation of network flows and command and control communication [44].

While ransomware have become increasingly sophisticated, traditional detection approaches struggle to detect evolving, polymorphic or zero-day ransomware, which can hide its behavior in multiple layers of execution. Further, each of these mentioned techniques has shortcomings. Some of these shortcomings may be overcome by packers and code obfuscation for static analysis, environmental checks or the use of delayed code execution for dynamic analysis or by encrypting the data for network analysis [9]. But the analysis of each of these techniques may be computationally intensive and solitary analyses may lead to biased interpretations causing the false positive and false negative values to escalate [11,12].

Machine Learning/Deep Learning has become a powerful tool for ransomware detection. Patterns extracted from binary files, dynamic traces, network flows can be used as input to ML/DL models to enhance the efficiency of ransomware detection. Most modern approaches depend on features derived from a single modality limiting robustness and generalization for ransomware analysis. A ransomware family might appear benign in a particular modality while showing malicious intent in another. This necessitates a multimodal approach that incorporates static, dynamic and network information. A technique can miss to see subtle patterns that indicate malicious intent if we rely on single modality.

It is also worth mentioning that despite the great progress in multimodal learning, ransomware detection remains fundamentally challenging due to high behavioral variability, polymorphism and adversarial evasion techniques. Models designed for detection as a static classification problem fail to adapt under distribution shifts across families or runtime environments. For instance, behaviorally related families such as LockBit are detected quite easily but highly polymorphic ransomware such as Dharma or WannaCry remains hard to detect.

Classic multimodal fusion techniques also assume uniform reliability across modalities. Therefore, those are vulnerable if certain modalities are unreliable or partially observable. This calls for an agentic ransomware detection where multiple specialized agents analyze different behavioral perspectives independently making a collaborative contribution toward a final decision. Agentic detection enables independent modality level reasoning, robust decision fusion, confidence aware abstention, and scalability for future threats. Using the implementation of agent level autonomy, the uncertainty associated with adversarial activities, sandbox evasion, network interruption and distribution shift can efficiently be dealt with meeting the real world operational challenges during ransomware defense. Finally, small LLM assisted feedback can enhance the detection process by providing contextual insights, synthesizing information across modalities, and suggesting improvements in feature integration [6]. This additional layer of intelligence ensures that the system can learn from complex patterns and improve generalization to previously unseen ransomware variants.

In this paper, the design of a multimodal multi agentic ransomware analysis (MMMA-RA) framework is proposed with the integration of static, dynamic and network modalities. Each modality acquires embedding with the help of auto encoders. The embedding is then combined by fusion so that the discriminative features of each modality are not lost. The classification is done using a family classifier based on transformer. At the end, an extra layer of intelligence is added in the form of AutoGen Multi agents based on Phi3.2B that can autonomously communicate and increase the confidence of final decision. MMMA-RA architecture focuses more on generalization capabilities and reliability rather than on forced classifications and decisions.

### 1.1. Main Contributions

Following is the main contribution of proposed MMMA-RA framework.

i. **Multi-agent multimodal ransomware classification framework:** In proposed MMMA-RA framework, a unified multi agent architecture for ransomware family classification is used that jointly use static, dynamic, and network modalities.

ii. **Class Imbalance Aware Multimodal Ransomware Classification:** MMMA-RA framework takes into consideration severe class imbalance by combining data balancing at train time and inversion frequency based class weighted optimization. This ensures that classification performance is not compromised while maintaining discriminative capability for both majority and minority classes.

iii. **Modality specific representation learning agents:** In MMMA-RA, independent modality agents based on autoencoders are designed to extract compact latent representations. It preserved modality specific semantics and avoided premature feature homogenization.

iv. **Contrastive regularized latent learning:** In MMMA-RA, a supervised contrastive learning objective is introduced. It is jointly optimized with reconstruction loss to minimize intra family variance and maximize inter family separation within each modality's latent space.

v. **Discriminative and aligned latent embedding:** In the proposed MMMA-RA framework, objective yields low dimensional, high information and class discriminative latent vectors $z_{\text{static}}, z_{\text{dynamic}}$ *and* $z_{\text{network}}$ forming a semantically consistent backbone for multimodal integration.

vi. **Calibrated Confidence Estimation for Operational Use:** For effective utilization of classifier in real world security applications, MMMA-RA employ post hoc probability calibration to ensure the classifier's outputs provide calibrated confidence estimates that better approximate the uncertainty of predictions.

vii. **Gated cross modality fusion mechanism:** MMMA-RA introduces a gated fusion strategy. It selectively integrates contrastively aligned modality embedding which enables effective cross modality interaction while mitigating noisy or redundant features.

viii. **Transformer based family level classification head:** MMMA-RA employs a classification head based on transformer to model inter-modality dependencies and perform accurate ransomware family level prediction.

ix. **Agentic Convergence and Improvement of Quality:** MMMA-RA shows the monotonic and stable convergence of the quality scores of the critic agent, the analyst agent and the composite quality scores of the three agents with absolute gains above +0.75 in 100 epochs for all agents and the composite quality scores approximate around 0.88 without any further optimization of the corresponding language models.

x. **Non intrusive agentic feedback mechanism:** In MMMA-RA, an agentic feedback module is presented that adaptively steers sampling and calibration during training. It improved minority family recall and convergence stability without modifying model weights.

## 2. Related Work

Ransomware has evolved into one of the most damaging classes of malware. It combines cryptographic primitives, stealthy execution and rapid monetization. Consequently, research on ransomware detection and classification has progressed from signature-based techniques to advanced machine learning and deep learning approaches that exploit static, dynamic and hybrid behavioral features [13,14]. This section reviews the existing literature with an emphasis on modalities, feature representations, learning models, datasets and other challenges.

Static analysis techniques extract discriminative features without executing the binary, including opcodes, API calls, byte sequences and PE header attributes. Santos et al. demonstrated that opcode sequences can effectively represent executables for unknown malware detection [7]. Raff et al. further advanced static learning by proposing end to end deep learning models that ingest entire PE files without handcrafted features [8]. Public datasets such as EMBER and BODMAS significantly accelerated static malware research by enabling large scale benchmarking [29]. Kim's NTMALDetect used native API system calls for malware detection [26], while Kuswanto and Anjad applied Random Forest and C4.5 classifiers on API call frequency features for ransomware detection [22]. Although computationally efficient, static approaches remain susceptible to obfuscation and packing techniques [11].

Initial ransomware and malware studies also focused on understanding attack workflows and behavioral characteristics. Andronio et al. proposed HelDroid, one of the earliest systems to dissect and detect mobile ransomware by analyzing application logic and user interface-driven behaviors [1]. Kharraz et al. provided a comprehensive examination of ransomware attack stages, highlighting encryption routines, file system interactions, and key management strategies [5]. These foundational works established that behavior-centric detection is more resilient than purely signature-based techniques.

Surveys on dynamic malware analysis by Egele et al. categorized analysis environments and common evasion techniques [27]. Afianian et al. in their paper emphasized the continuous arms race between malware authors and defenders [28]. Connolly and Wall later introduced a taxonomy of crypto ransomware countermeasures, organizing detection, prevention, and recovery strategies within a unified framework [4]. Dynamic analysis captures runtime behaviors such as system calls, file operations, registry modifications, and network activity. Cen et al. proposed RansoGuard, an RNN based framework that focuses on pre-attack sensitive APIs to enable early ransomware detection before encryption is completed [15]. Molina et al. explored ransomware family attribution using pre-attack "paranoia" activities [21]. Authors demonstrate that early behavioral signals can effectively distinguish ransomware families.

Process and memory-based analysis has also emerged as a robust detection direction. Koyirar et al. investigated ransomware detection through process memory inspection, improving resilience against code obfuscation [18]. Network-level behavioral modeling using flow-based tools such as CICFlowMeter further supports ransomware and anomaly detection in distributed environments [32]. Recent ransomware research increasingly relies on deep learning particularly sequence models and transformers. Karbab et al. introduced SwiftR, a cross-platform ransomware fingerprinting framework using hierarchical neural networks trained on hybrid static and dynamic features [19]. Zahoora et al. proposed a deep contractive auto encoder combined with ensemble classifiers for zero-day ransomware detection. Authors significantly addressed generalization challenges in [20].

Transformer architectures have demonstrated strong performance in modeling long range dependencies. Gaber et al. proposed Pulse, which applies transformer based function classification over assembly level representations for zero-day ransomware detection [14]. Alzahrani et al. introduced RansomFormer, a cross modal transformer architecture that fuses byte level and API call features. The authors achieved superior detection performance and robustness compared to unimodal approaches [13]. Explainability has become increasingly important for operational deployment of ransomware detection systems. Gulmez et al. proposed XRan, an explainable deep learning framework using dynamic analysis to provide interpretable detection decisions [16]. Recent work on structured heat map learning using CAPEv2 further demonstrated the feasibility of explainable multifamily malware classification [49]. Hybrid detection systems combining multiple feature spaces and learning paradigms have also shown promising results. BN and Brahmananda introduced a multiclass hybrid ML framework augmented with semantic similarity for ransomware detection and criticality assessment [17]. Al-Rimy et al. proposed an early ransomware detection model using incremental bagging with enhanced semi random subspace selection, achieving improved performance during early execution stages [23]. The effectiveness of ransomware detection approaches strongly depends on dataset quality and diversity. Public datasets such as EMBER [29], BODMAS [30], CIC-AndMal-2020[31] and CICFlow based traffic datasets [32] are widely used. However, many recent studies still rely on private or custom datasets due to the lack of continuously updated and labeled ransomware datasets. This limits reproducibility and makes fair quantitative comparison across studies challenging. In order to provide a clearer empirical perspective, Table 1 presents a quantitative comparison of representative ransomware detection approaches from the recent literature (2019–2025), focusing on learning models, feature modalities, detection scope, datasets and reported performance metrics. The comparison highlights that while transformer and deep learning based approaches often achieve high detection accuracy, their effectiveness is typically constrained to specific feature modalities or detection scopes. Moreover, most existing solutions rely on a single analytical perspective. It reinforces the need for more holistic detection paradigms. Despite significant advances, several challenges, including robustness against obfuscation, packing and adversarial manipulation [23], reliable detection of zero day and low frequency ransomware families under severe class imbalance [20] and early detection before irreversible encryption damage occurs [15] remain unresolved.

It is worth mentioning that recent advances in multi agentic AI enabled autonomous agents to collaborate, share insights and adapt to complex environments in real time [25]. In ransomware analysis, this approach is particularly promising for ransomware detection where traditional methods often fail against evolving variants. A multi agent framework can assign agents to monitor files, processes and network activity simultaneously, sharing observations to identify ransomware families accurately. By analyzing behavioral patterns collectively such systems offer a proactive and adaptive defense against emerging ransomware threats. As shown by S. K. Fatima et al. in [40], AutoGen enables the seamless coordination and communication between multiple intelligent agents. It improves scalability and accuracy in complex analytical workflows. AutoGen can be used to iterative data processing and predictive modeling. It results in more rich insights from ransomware sample datasets.

The reviewed literature suggests that no single feature representation or learning model is sufficient to address the evolving landscape of ransomware threats. Static methods offer efficiency but lack robustness, dynamic approaches provide richer behavioral insight but incur runtime overhead and deep learning models often struggle with interpretability and generalization when used in isolation. Therefore, multi model multi agent ransomware detection framework is proposed as it is a requirement of the time. A system can collaboratively detect, classify and attribute ransomware attacks with higher robustness and resilience by using multiple agents. The multi agent architecture aligns with Artificial General Intelligence paradigms by enabling decentralized cognition, cooperative reasoning and adaptive decision making. It demonstrates that AGI inspired systems can be applied to complex cybersecurity tasks [10]. Our proposed system does not constitute full AGI. Rather, it embodies foundational AGI principles within a constrained ransomware domain. Each agent specializes in distinct feature spaces and learning paradigms. Multi-agent coordination enables early warning, cross validation of decisions and adaptive in ransomware domain. Each agent specializes in distinct feature spaces and learning paradigms. Multi-agent coordination enables early warning, cross validation of decisions and adaptive response. Further, multi-model fusion mitigates individual model weaknesses. MMMA-RA framework aligns with the growing complexity, heterogeneity and adversarial nature of modern ransomware. It also directly addresses the limitations identified in existing single model approaches.

## 3. Methodology

We propose building a single pipeline that trains modality encoders, aligns and fuses latent spaces, learns a family classifier with class weighted Contrastive Auto encoder, applies post hoc calibration and runs an Auto Gen multi agent loop to steer sampling and inference thresholds. Figure 2 shows the overall architecture of proposed MMMA-RA framework.

In the proposed MMMA-RA the complete flow of machine learning mindset i.e. data collection, data preprocessing, feature engineering, model training and optimization, and model evaluation is adopted at discussed by A. Khan et.al in [42]. The pipeline starts with the processing of the raw input modalities. To begin with, each unique input modality is trained on individual modality encoders. The individual modal encoders have specialized roles in attempting to derive meaningful representations of different forms of data. After this is accomplished, a very important step in aligning these derived representations is followed. In this stage, it is ensured that different modal representations are uniform. Our final step for multi model representation is to incorporate a fusion technique where all these representations will finally be put into a complete vector. Once a comprehensive multimodal representation is generated, it is used to train a basic classifier model. A family classifier is learned using a direct representation on a fusion latent vector. The learning objective of this classifier contains a dedicated part which is a class weighted Contrastive Auto-Encoder. The addition of a class weighted CAE is important since it accounts for a problem of class imbalance which can arise during learning. A class weighted CAE gives a higher priority to classes observed less during learning. This gives a higher boost to a class classifier in learning robust models for challenging classes. Once the basic training is over, the entire system is further improved by using approaches which are meant for improving both the output confidence and the inference step. As a first step towards this goal, post-hoc calibration is performed on the output confidence of the classifier. This is a critical step in ensuring that the output confidence of a predictive model is a good indication of the real confidence level of the predicted output being correct. At the end, the system integrates an Auto Gen multi agent loop. The entire loop consists of a series of interlinked agents. These agents include User Proxy, a Critic and an Assistant agent. This proposed multi agent system serves a very important purpose in providing an optimized mechanism that can be used to improve the performance of the system. Detailed descriptions of individual components are given in subsequent sub-sections. this goal, post-hoc calibration is performed on the output confidence of the classifier. This is a critical step in ensuring that the output confidence of a predictive model is a good indication of the real confidence level of the predicted output being correct. At the end, the system integrates an Auto Gen multi agent loop. The entire loop consists of a series of interlinked agents. These agents include User Proxy, a Critic and an Assistant agent. This proposed multi agent system serves a very important purpose in providing an optimized mechanism that can be used to improve the performance of the system. Detailed descriptions of individual components are given in subsequent sub-sections.

| Ref | Year | Method / Model | Feature Type | Detection Scope | Dataset | Performance |
|---|---|---|---|---|---|---|
| Multi Model Multi Agent Ransomware Analysis [Proposed] | 2026 | MMMA-RA Multi Model Multi-agent framework | Static, Dynamic, Network | Detection, family classification, zero-day detection | Custom | Accuracy: 95.9%, F1: 0.946 |
| RansomFormer: A Cross Model Transformer Architecture [13] | 2025 | Transformer (cross-modal) | Byte-level + API | Detection + family classification | Private + public datasets | Accuracy: 97.8%, F1: 0.977 |
| Zero-day ransomware detection with Pulse [14] | 2025 | Transformer | Assembly functions | Zero-day detection | Custom ransomware binaries | AUC ≈ 0.99, F1:0.96 |
| RansoGuard: A RNN-based framework [15] | 2025 | RNN (LSTM/GRU) | Pre-attack APIs | Early detection | Windows ransomware samples | Accuracy: 96.4% |
| XRan: Explainable ransomware detection [16] | 2024 | CNN + XAI | Dynamic API traces | Detection | CIC ransomware datasets | Accuracy: 98.1%, F1: 0.98 |
| Revolutionizing ransomware detection [17] | 2024 | Hybrid ML | API + semantic features | Multiclass detection | Mixed families | Accuracy: 97.2%, F1: 0.96 |
| Efficient ransomware detection through process memory [18] | 2024 | ML classifiers | Process memory | Detection | OS memory dumps | Accuracy: 95.6% |
| SwiftR: Cross-platform ransomware fingerprinting [19] | 2024 | RF, SVM XGBoost | Static + API | Detection | Public datasets | Accuracy: 96–98% |
| Zero-day ransomware attack detection [20] | 2023 | Hierarchical NN | Static + Dynamic | Detection + fingerprinting | Cross-Platform Samples | Accuracy: 98.6% |
| On ransomware family attribution [21] | 2022 | Autoencoder + Ensemble | Behavioral | Zero-day detection | Custom | Accuracy: 97.4% |
| Ransomware detection based on API call frequency [22] | 2021 | ML classifiers | Pre-attack behavior | Family attribution | Network & host data | Accuracy: 94.8% |
| Crypto-ransomware early detection [23] | 2021 | RF + C4.5 | API frequency | Detection | Windows samples | Accuracy: 93.7% |
| Ransomware detection using ML [24] | 2019 | Ensemble ML (incremental bagging) | Behavioral | Early detection | Large ransomware datasets | Accuracy: 97.1% |

**Table 1: Quantitative comparison of recent ransomware detection and classification approaches**

### 3.1. Input Layer

The input layer contains data from three different modalities comprising static, dynamic and network features. Static features consist of PE headers, opcode n-grams, entropy, imports and exports. Dynamic features consist of API call traces, registry and file system activity and process level behavior. Lastly Network features consist of flow duration, packet statistics, protocol usage and timing characteristics.

The system detects ransomware data in three heterogeneous forms that is static, dynamic and network using particular modality specific deep contrastive auto encoders as shown in Figure 3. These are trained to obtain powerful and discriminative latent presentations. The input to each modality is of the form of a feature matrix with rows representing samples and columns representing modality centric attributes. The static inputs are those which capture PE structure signals such as headers, opcode n-grams, entropy values and imports/exports. The dynamic inputs are those which capture temporal execution behavior such as API sequences and registry reads, writes and deletions, file system activity such as encryption bursts and process level actions. The network inputs are those that capture communication behavior such as flow statistics including duration, lifetime and throughput etc. During inference if a modality is not available, this will be masked in the corresponding vector in the latent space. This vector will not be considered in the fusion stage because of the gated fusion strategy since it allows graceful degradation.

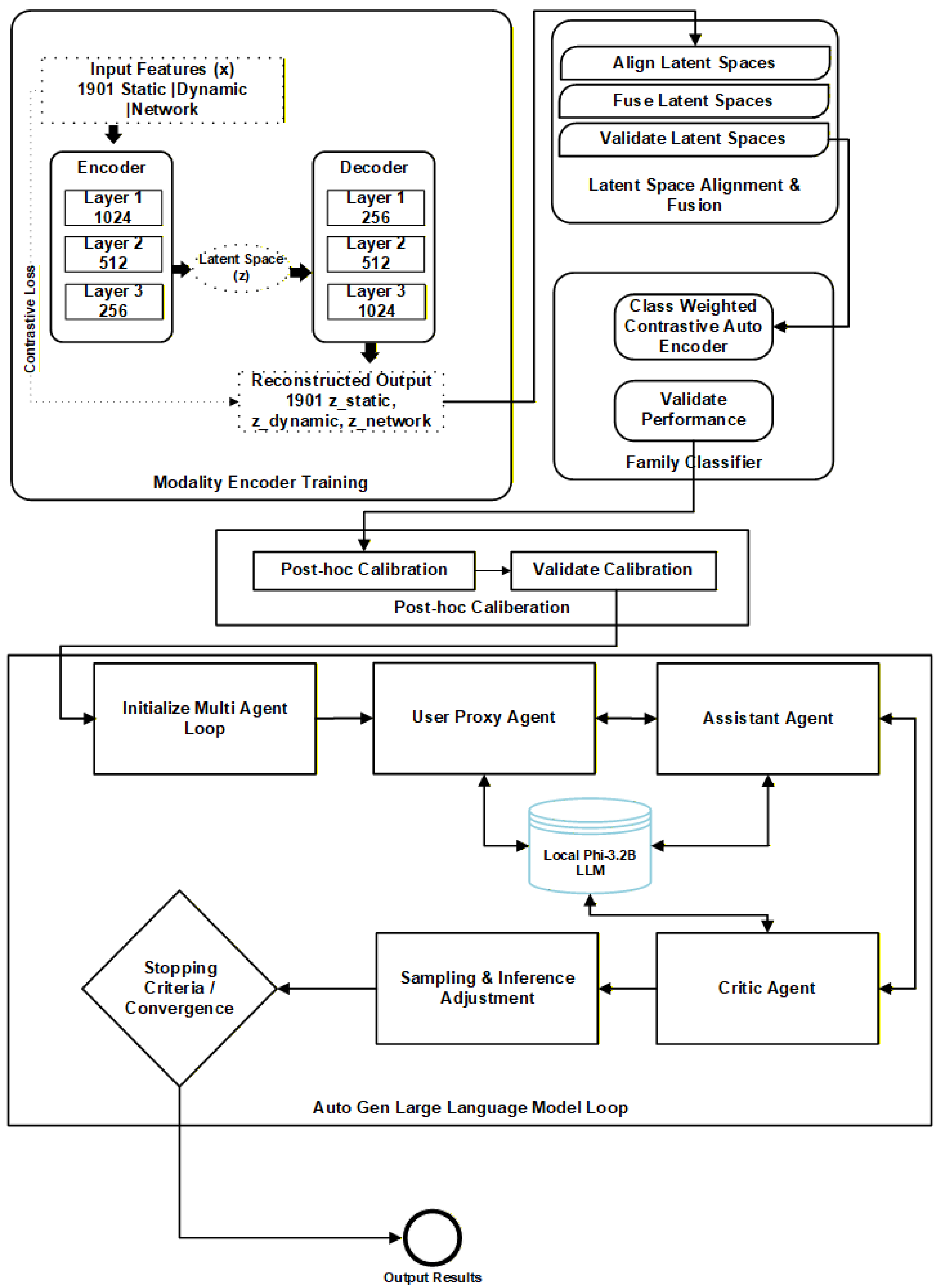

**Figure 2: Multi Model Multi Agent Ransomware Analysis (MMMA-RA) – Proposed Methodology**

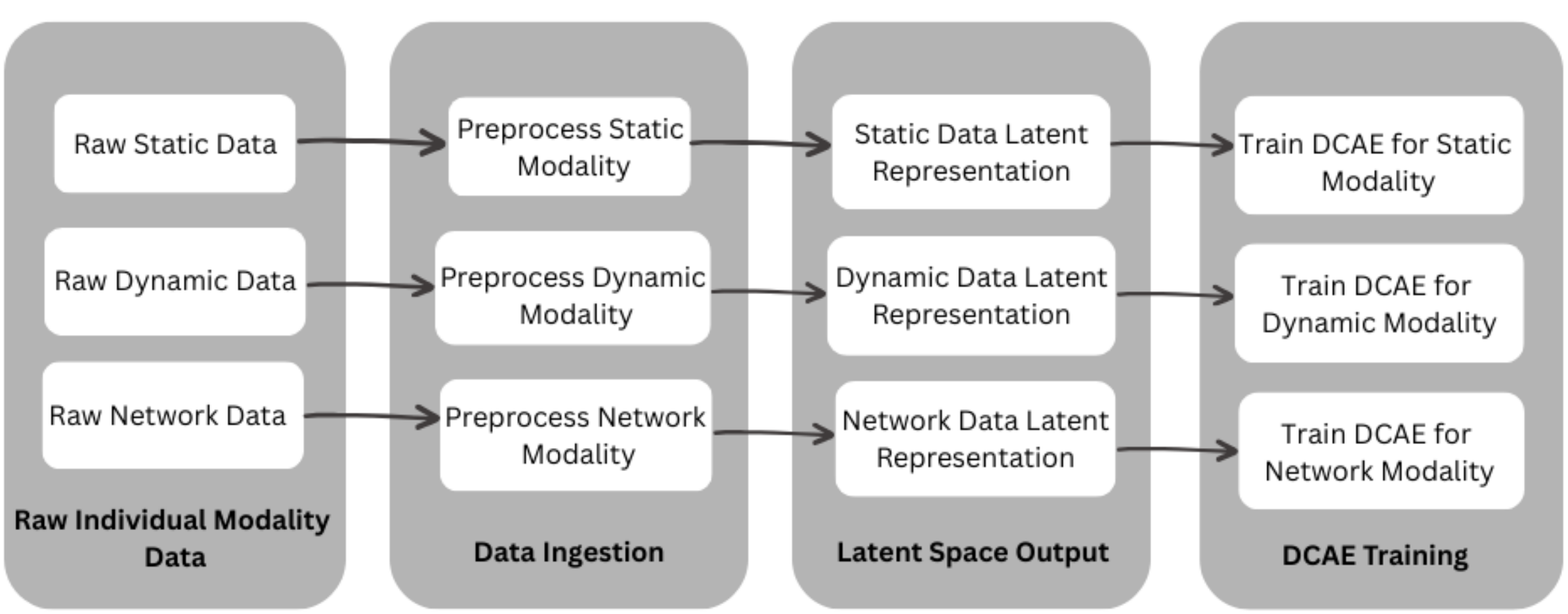

**Figure 3: Input Layer & Application of modality specific DCAE**

### 3.2. Modality Specific Deep Contrastive Auto Encoders (DCAEs)

The Static DCAE utilizes a four (4) layer encoder (1901 → 1024 → 512 → 256 → 128) and a symmetric decoder. Khan et al [41] discussed that deep contrastive autoencoders combine contrastive objectives with latent space reconstruction to enable robust feature extraction for complex data distributions. Similarly, dynamic and network DCAEs use analogous architectures tailored to their respective feature dimensions. Each decoder attempts to reconstruct the original input $\hat{x}_m$ and the reconstruction loss ensures preservation of critical modality information. In parallel, a supervised contrastive loss encourages embedding from the same ransomware family to cluster together while pushing apart embedding from different families. The combined objective as shown in Eq. (1) reduces intra family variance and strengthens inter family separation.

$$L_{AE} = \| x - \hat{x} \|^2 + \lambda L_{\sup} \quad (1)$$

In equation (1), $L_{AE}$ denotes the total autoencoders loss where $\|x-\hat{x}\|^2$ is the reconstruction loss between the input $x$ and its reconstruction. $Lsup$ represents the supervised loss component while λ is a weighting parameter that balances reconstruction accuracy and supervised learning.

It produces modality specific latent vectors $z_{\text{static}}, z_{\text{dynamic}}, z_{\text{network}}$. These embeddings are low dimensional, high information and class discriminant representations. These serve as the aligned backbone for multimodal later fusion and classification at the family level. Overall architecture of deep contrastive auto encode is depicted in Figure 4.

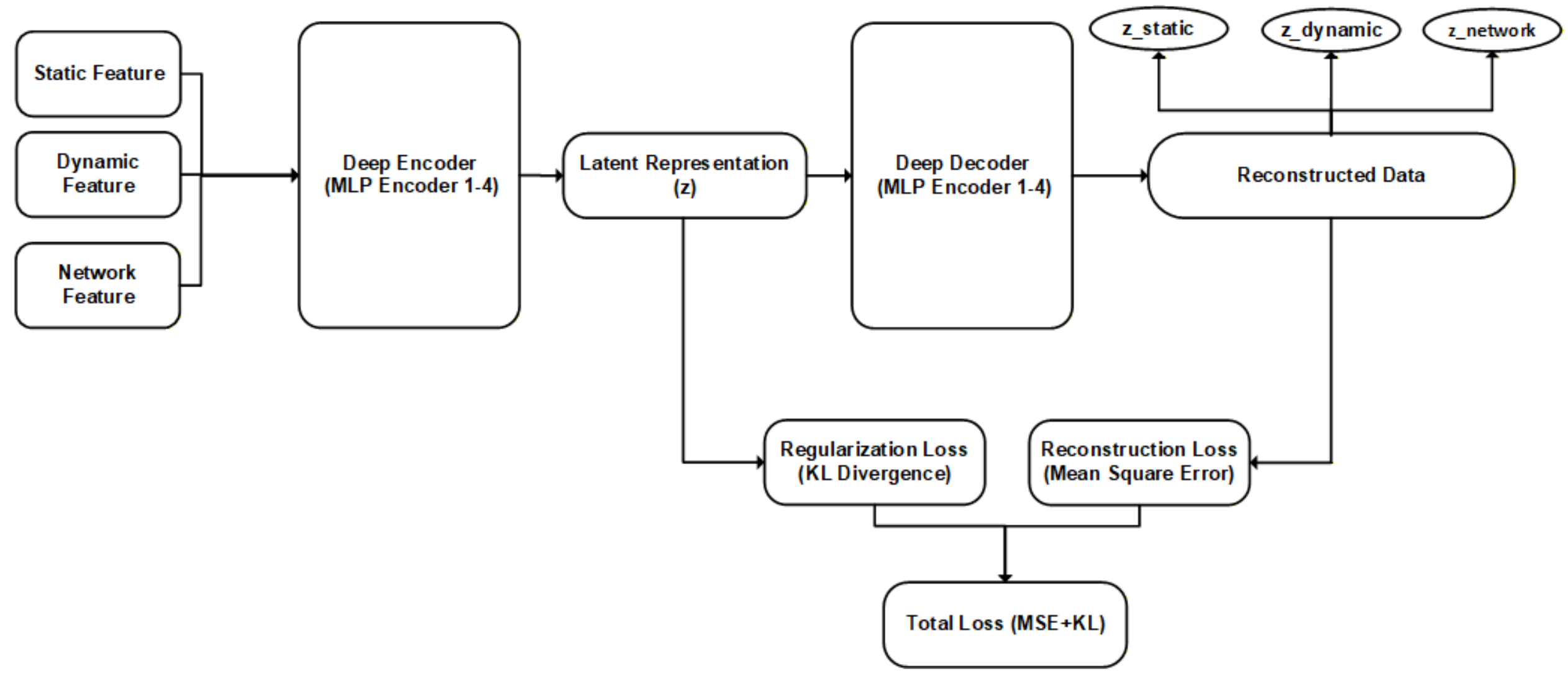

**Figure 4: Architecture of Contrastive Auto encoder**

Each modality is encoded into a latent representation using a four (4) layer encoder and a symmetric decoder. Both encoder and decoder are trained jointly with reconstruction and supervised contrastive losses. The output latent vectors are $z_{\text{static}}$, $z_{\text{dynamic}}$, $z_{\text{network}}$. Traditional autoencoders (AE) optimize only a reconstruction loss.

This learns features that best reconstruct the input. However, reconstruction quality does not imply useful classification features. It is observed that:

- ransomware embedding show that reconstructed features may capture low level noise rather than semantics useful for classification.
- In cybersecurity tasks, features that minimize reconstruction loss alone can be overly focused on frequent patterns and fail to capture subtle malicious signals.

Therefore, while auto encoders are effective as denoisers or generative models, they are not guaranteed to produce features with strong class separability. Contrastive learning methods specifically contrastive auto encoders optimize representations such that similar samples are closer and dissimilar ones farther apart in latent space. Contrastive representations have been shown to outperform purely reconstructed embedding for classification tasks especially when multiple modalities are involved.

- Chen et al. [36] demonstrated that contrastive loss produces embedding with better class separability compared to traditional auto encoders.
- Misra & van der [37] found that contrastive learning yields clusters that align well with semantic classes without supervision.

In malware/ransomware specifically:

- Razi Mahmood et al. [38] applied contrastive learning to behavioral logs, showing that contrastive embedding improves classification accuracy over baseline auto encoders.

Thus, if an auto encoder is used it should be contrastive or supervised in nature rather purely generative.

### 3.3. Fusion Module

A unified multimodal latent vector $z_{\text{fused}}$ is produced using the concatenation of vectors. Multimodal fusion begins with the process of matching the latent representations of each modality, i.e. $z_{\text{static}}$, $z_{\text{dynamic}}$, and $z_{\text{network}}$. In order to bring the three modalities to a similar sample, the alignment stage is required to perform the strict label matching and family level balancing. This is done by up sampling to eliminate the imbalance in each modality. The Fusion Module reforms concatenation fusion of the three latent vectors after the alignment stage to form a single representation. This representation is described as $z_{\text{fused}} = [\, z_{\text{static}} \parallel z_{\text{dynamic}} \parallel z_{\text{network}} \,]$. Resultantly, this fused embedding captures complementary structural, behavioral and network level semantics in a single and enriched feature vector ready for downstream family classification.

### 3.4. Family Classifier

Finally, a fully connected neural classifier trained with the fused latent representation produces the final ransomware family prediction. This classifier performs inverse frequency class weighting to account for severe family imbalance and applies post hoc probability calibration to improve the reliability of the classifier's confidence estimate.

The Family Classifier is implemented as a multi-layer perceptron that operates on the fused latent representation produced by the concatenation of the static, dynamic and network embedding as shown in Figure 5.

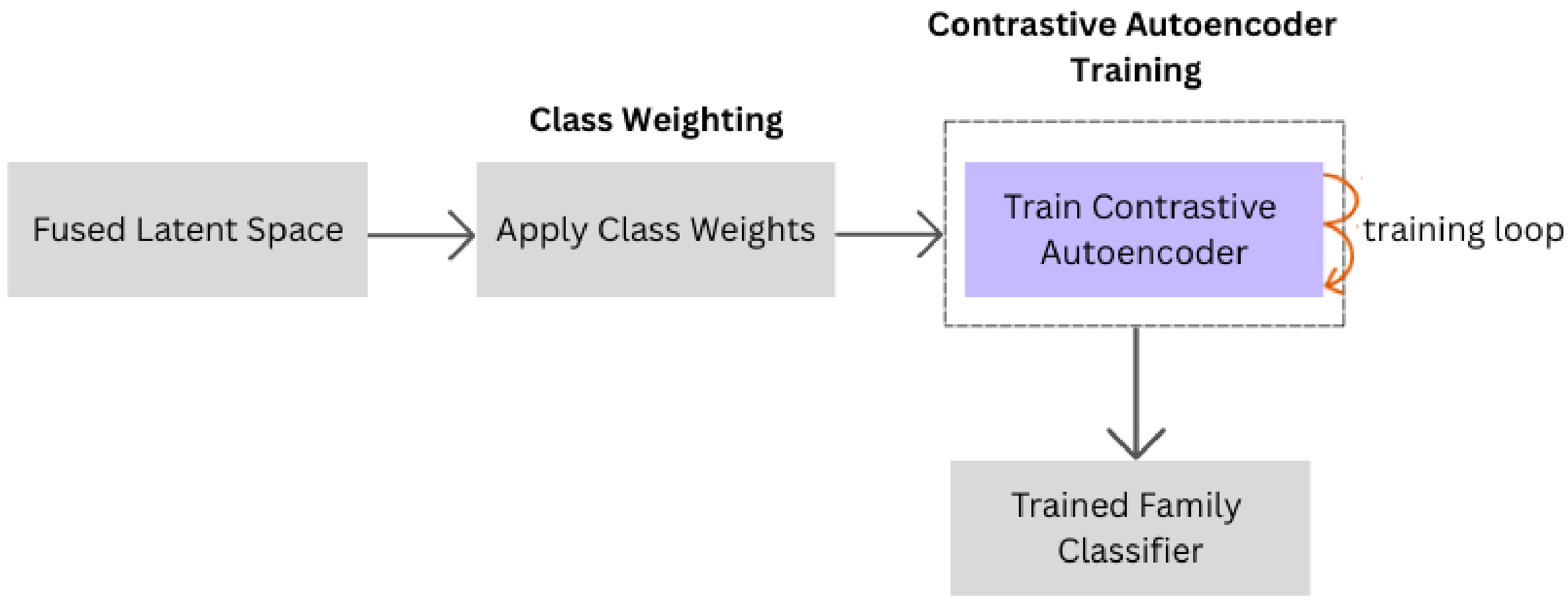

**Figure 5: Training pipeline for contrastive auto encoder based family classification**

Formally, the classifier input is

$$Z_{fused} = Z_{static} \| Z_{dynamic} \| Z_{network} \quad (2)$$

In (2), $Z_{static}$, $Z_{dynamic}$ and $Z_{network}$ denote the latent representations extracted from static, dynamic and network modalities respectively. The operator ‖ indicates feature concatenation resulting in the fused representation $Z_{fused}$ which captures complementary information across modalities.

In a single cohesive latent space, this fused vector combines the disparate behavioral cues discovered through PE structure, execution behavior and network flow semantics. The fused representation has a discriminative structure. It can be used directly by a downstream classifier without the need for extra normalization or balancing because each modality has already been contrastively aligned during training.

The multi-layer perceptron processes this input through multiple fully connected layers. Each layer applies linear transformations followed by nonlinear activation functions. This hierarchical transformation can be expressed as

$$h_1 = \sigma\left(W_1 z_{\text{fused}} + b_1\right), h_2 = \sigma\left(W_2 h_1 + b_2\right), \ldots \tag{3}$$

where $\sigma(\cdot)$ denotes a nonlinearity such as ReLU and each layer progressively extracts higher level family discriminative features. We addressed the significant imbalance across ransomwares family problem by inverse frequency class weights $w_c$are applied during optimization. This ensures that minority families exert proportionally stronger influence on the classifier. This improves robustness and generalization across all families.

In the output stage, the final layer produces a probability distribution over all ransomware families using the softmax function:

$$y_c = \frac{exp\left(W_c h_k + b_c\right)}{\sum_{j=1}^{C} exp\left(W_j h_k + b_j\right)} \tag{4}$$

where $C$ is the number of families, $c$ is the class index e.g {c € 1,2,…., C} and $h_k$is the final hidden representation. This probability vector reflects how strongly the classifier associates the input sample with each possible ransomware family.

The predicted family is derived simply as the class with the maximal probability

$$\hat{y} = \arg\max_{c} y_c \tag{5}$$

The model’s confidence is defined as the probability assigned to the predicted class

$$\text{Conf}\left(\hat{y}\right) = \max_{c} y_c \tag{6}$$

### 3.5. Post-hoc Probability Calibration

Despite normalized probabilities from softmax outputs, it has been observed that current models can produce overconfident outputs. In that regard, vector scaling techniques have been employed to improve its probability calibration and address overconfident outputs as shown in Figure 6. The post calibration process is done after training and doesn’t involve adjusting the parameters.

The calibrated output of probability encoded in the model above produces confidence estimates of predictions that better represent actual predictive uncertainties. Low confidence predictions tend to correspond to samples with overlapping or ambiguous behavioral characteristics, while high-confidence predictions tend to correspond to samples for which there exists strong and consistent evidence across static, dynamic and network attributes.

In practice, this means that in addition to determining which ransomware family is most likely the classifier also offers an interpretable confidence estimate that measures how certain its prediction is. While lower confidence values indicate ambiguous or partially overlapping behavioral traits, high confidence values generally indicate strong and modality consistent evidence across static, dynamic and network features. For operational cybersecurity settings, this calibrated probability output is essential because it allows analysts and automated response systems to rank alerts according to both estimated reliability and predicted family.

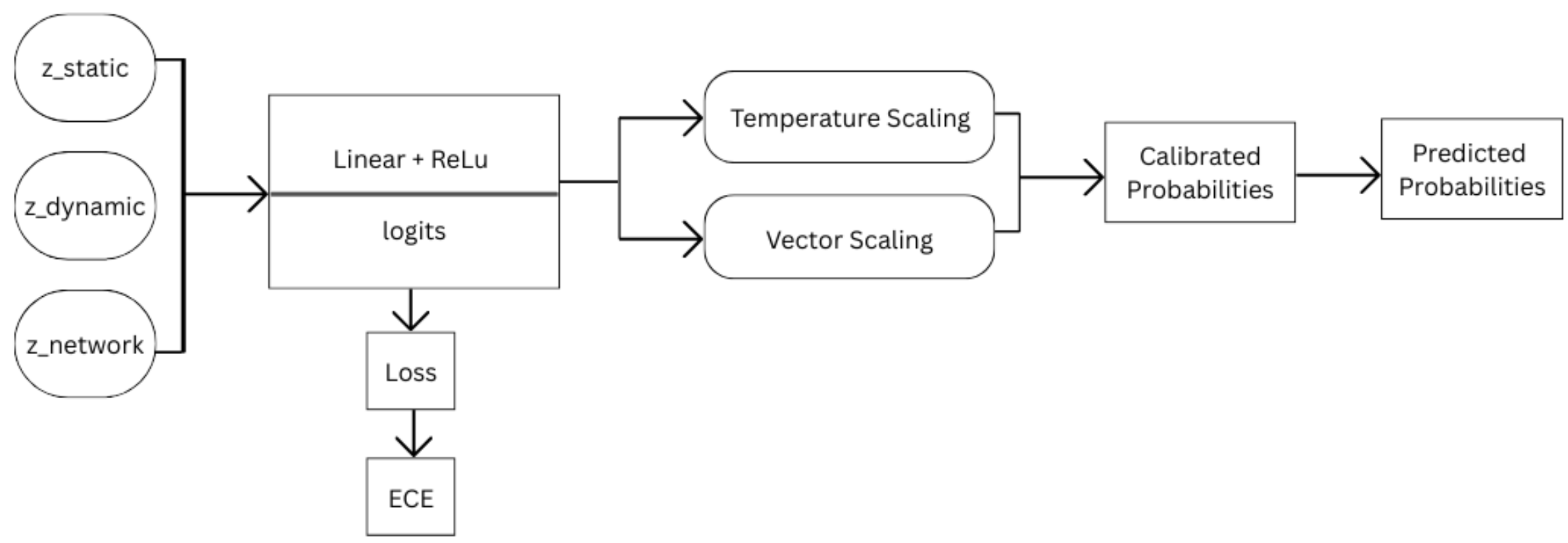

**Figure 6: Post Hoc Calibration**

### 3.6. AutoGen Large Language Model Loop

Three lightweight agents Analyst, Feedback and Predictor form a feedback loop that adjusts sampling policies, thresholds and hyper parameters without modifying model weights as shown in Figure 7.

Three cooperative agents User Proxy Agent, Critic Agent and Assistance Agent are instantiated using a local lightweight Large Language Model (Local Phi3.2B Model). This comprises the lightweight yet fully automated multi agent orchestration layer included in this work. These agents work in a structured feedback loop and exchange brief role conditioned messages at each training epoch. They assess predictions, identify weak ransomware families and affect adaptive rebalancing, sampling and calibration mechanisms. The critic agent is a large language model-based evaluator rather than a reinforcement learning critic because the mechanism is fully LLM driven. It does not make use of a reinforcement learning policy network. All agents interact via deterministic and template controlled prompts to guarantee uniform structure. The agents produce low variance outputs with high clarity.

### 3.6.1. User proxy Agent

The current epoch's aggregate model statistics are sent to the User Proxy Agent. Its function is to generate a concise triage message in a predetermined format. It includes confidence profile analysis, ransomware family prediction with confidence score and a suggested next step. Instead of altering the classifier, the agent generates structured text that the Critic Agent then assesses for accuracy and completeness.

### 3.6.2. Critic Agent

The Critic Agent is not a reinforcement learning value function. It is implemented as a Large Language Model evaluator. The model is compelled by its system prompt to produce precisely four fields. These four fields are Guardrail, Missing Element, Strength and Flaw. The critic evaluates the quality of the Analyst Agent's reasoning. It then finds the context that is missing. Finally, it decides whether the model's confidence level justifies escalation. The critic completes a second task in the critic_plus strategy. It is prompted with a specific query, "name the weak families," and the model extracts low performing ransomware families from natural language text based on F1 score per class scores, ECE and epoch specific performance drops. These families are parsed by the code using _extract_target_families(), creating a set of oversample_targets. In the upcoming epoch, these targets have a direct impact on hard mini-batch selection, adaptive oversampling and pseudo label fine tuning. The Critic Agent functions as an LLM based performance diagnostician as a result. Its natural language feedback is transformed into structured control signals (target families, reliability scores, oversampling weights) rather than computing numerical rewards. A reliability score (critic_reliab) is updated every epoch based on the change in F1 score of previously targeted families. This creates a soft trust metric for the critic.

### 3.6.3. Assistance Agent

The Assistance Agent generates a short forecast statement about expected performance trends. It uses the same statistics available to the Critic Agent but produces forward looking risk assessments. Although this agent does not alter training dynamics directly, its dialogue is logged and contributes to the clarity, jargon and compliance composite metric computed per epoch. This composite score is used in the final analysis to quantify emergent coherence of the agentic subsystem. A sample interaction dialogue between agents is shown in Figure 8.

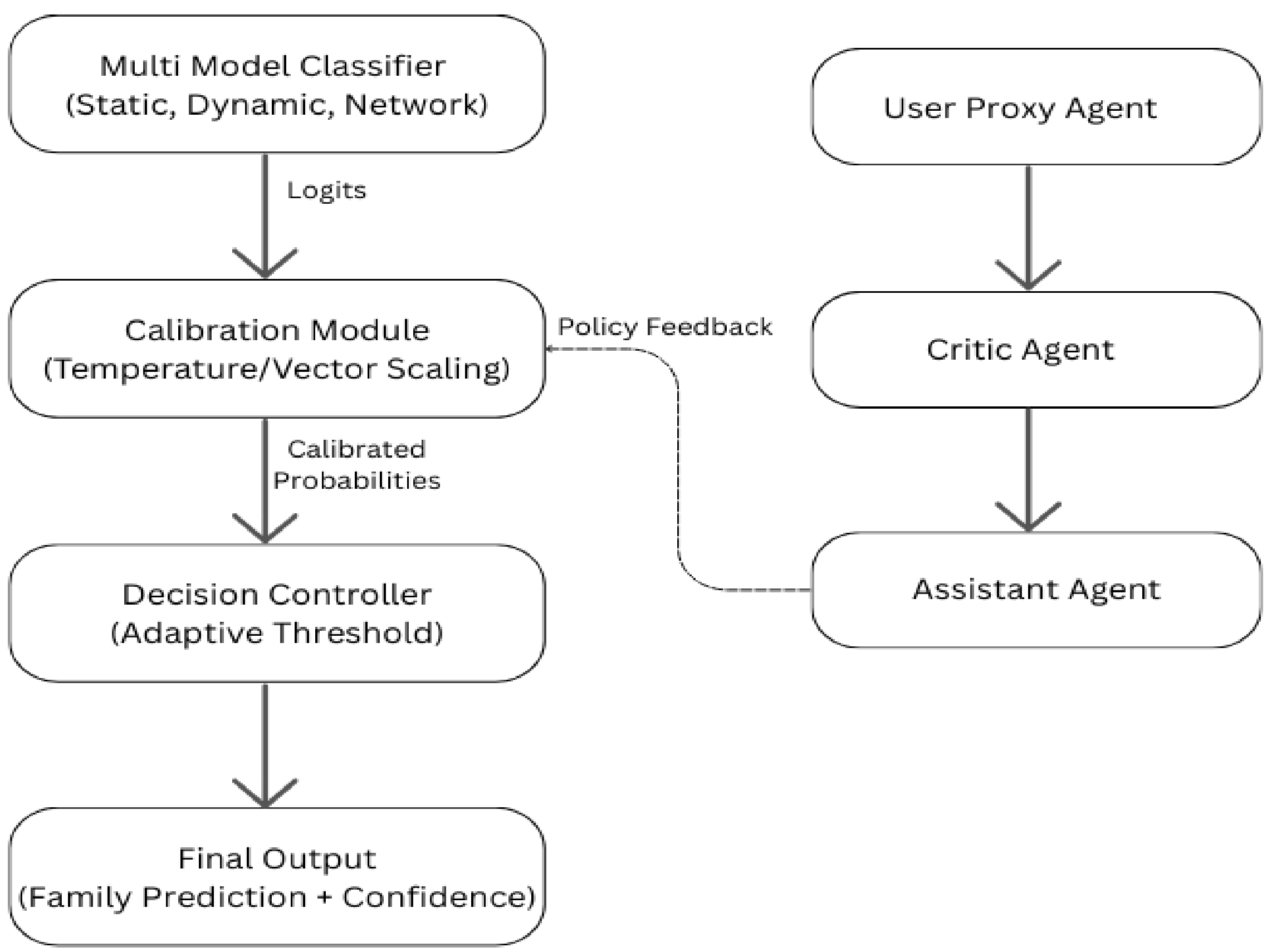

**Figure 7: Integration of AutoGen Agentic Loop**

### 3.6.4. Agent Interaction Cycle

Every epoch triggers the following interaction sequence.

i. Model inference summary is given to User Proxy Agent.
ii. User Proxy Agent outputs a structured diagnosis and a recommended next step.
iii. Assistance Agent output is forwarded to Critic Agent.

```
AnalystAgent:
Analysis: Ransomware with high similarity to known families.
Prediction: 0 | Confidence: 99.5%
Next step: Review the ransomware's encryption method and payload for further analysis.

FeedbackAgent:
Flaw: The ECE (Expected Calibration Error) is relatively high at 0.055, indicating that the model's confidence in its predictions does not
Strength: The top-1 accuracy is very high at 91.1%, showing that the model is performing well in predicting the correct class for the majo
Missing Element: The analysis lacks information on the model's performance across different classes or subgroups, which is crucial for und
Guardrail: if top-1 < 55% or margin < 10%, escalate.

PredictorAgent:
Analysis: The F1 score after the model's adjustments has significantly improved, indicating a better balance between precision and recall.
Prediction: The model is likely to perform well in predicting family or risk, with a high degree of accuracy and reliability.
Note: The model's performance has improved, but further testing and validation are recommended to ensure its robustness across different d

AnalystAgent:
Analysis: Ransomware sample exhibits high similarity to known families with a top-1 score of 84.8%.
Prediction: 0 | Confidence: 98.6%
Next step: Continue with automated analysis and monitoring.

FeedbackAgent:
Flaw: The average margin is below the guardrail threshold of 10%.
Strength: The ECE (Expected Calibration Error) is well below the guardrail threshold of 0.1, indicating good calibration.
Missing Element: The top-1 accuracy is not provided, which is necessary to fully assess the model's performance.
Guardrail: if top-1 < 55% or margin < 10%, escalate.
```

**Figure 8: Sample Dialogue**

iv. The critic evaluates flaws, missing elements and flags escalation cases. The second critic pass (critic_plus strategy) receives current F1 scores per family, ECE/accuracy trends and list of lowest F1 families. It returns textual names of families needing improvement.

v. The system parses critic responses: $\text{targets} = \text{ExtractWeakFamilies}(\text{critic_reply})$. These families become entries in oversample_targets.

vi. Target families receive amplified sampling probability.

vii. Assistance Agent summaries are written to disk and later contribute to clarity and composite quality metrics.

This three agent loop introduces structured reasoning into the training process while avoiding any direct modification of model weights by the LLMs. Instead, the LLM subsystem acts as a meta controller that shapes sampling, rebalancing, confidence fallback and calibration selection.

# 4. Experimental Setup

## 4.1. Dataset Overview

Ransomware behavior could be detected in the following three attack modalities.

i. Static: PE headers/sections/flags, imports/exports, opcode n-grams, raw bytes/entropy

ii. Dynamic: API/system call, registry/file system activity, process behavior

iii. Network: DNS/TCP flows arrival, flow size/rates, protocols, inter-arrival time, TCP flags/window sizes

The hashes of groups are required in order to ensure non-cross-modality leakage of rows of a given sample in modalities.

#### 4.1.1. Dataset Construction and Sample Acquisition

The project utilizes a public dataset of ransomware. This data set is available on Zenodo [33]. The dataset has labeled samples of different types of ransomware as well as benign samples. However, the six classes of great importance are focused from the data set. These six classes are Benign, Ryuk, LockBit, Dharma, Shade and WannaCry.

The original binary files were obtained using VirusTotal by hashing the samples using the VirusTotal platform. The samples had to be successful at being acquired with the ability to be executed within a sandbox environment. This ensured a total of 500 samples per class amounting to 3000 samples of binaries after consideration of class balancing.

#### 4.1.2. Static Data Set (Zenodo Dataset)

Table 2 list important properties of Zenodo dataset. Further, a brief overview of features is also given in this sub section.

**Table 2: Static Data Set**

| Total Samples | No. of Features | No. of Classes | Benign | Ransomware |
|---|---|---|---|---|
| **21,752** | 1,901 | 12 | 10,876 | 10,876 |

i. The indications of packing include PE metadata, PE layout PE filesize, PE imagebase, PE entry point, PE numsections, PE loaders PE payload layout variations.
ii. Embedded payloads / config flagged options Flagged options: hasdebug, hastls, hasresources, early logic hiding, embedded payloads / config.
iii. Imports/Exports: API is dynamically resolved because the value of imports, the dllcount and the minimum constant imports are all dynamic.
iv. The statistic of entropy: entropy avg/max section, packing/ compression/ crypto.
v. Opcodes: one/three gram unpack/ decrypt idioms, family prints.
vi. Labels: Category, Family.

### 4.1.3. Dynamic Analysis and Feature Extraction Using CAPE v2

#### 4.1.3.1. Sandbox Execution Environment

The dynamic behavioral analysis was conducted by using CAPE-V2. CAPE-V2 is an advanced malware sandbox that can be used for automated analysis. This was done by executing the ransomware sample inside a Windows virtual machine that was isolated to track system level events [34].

In the sandbox environment, there are detailed behavioral events recorded related to API call activities, process creation and interactions with the Windows Registry and file systems. Execution is done within a fixed time window to allow adequate time to observe activities related to encoding routines and modification of systems by ransomware.

#### 4.1.3.2. Dynamic Feature Extraction and CSV Generation

For each executable that has been executed, the CAPE v2 has produced the behavior reports as structured json. These were processed for the extraction of the dynamic behavior features that were then processed to generate CSV. Each sample's dynamic behavior was converted into a fixed length numerical feature vector and stored as a row in the dynamic feature CSV dataset. Table 3 list important properties of dynamic dataset. Further, a brief overview of features is also given in this sub section.

**Table 3: Dynamic Dataset**

| Total Samples | No. of Features | No. of Classes | Benign | Ransomware |
|---|---|---|---|---|
| **3000** | 77 | 5 | 500 | 2500 |

i. Process behavior: malicious/ suspicious/ under observation, subsurface execution to encrypt/ disable tools.

ii. File I/O: high speed/ large scale encryption tendencies, malicious/suspicious/text/ unknown.

iii. Dynamic crypto, FS, networking resolution dubbed controversial API/DLL.

iv. Classes: Class, family, category.

### 4.1.4. Network Traffic Collection and Feature Extraction

#### 4.1.4.1. Network Traffic Capture

When the CAPE v2 sandbox was dynamically executing the network communications that resulted from the execution of each of these files were captured. The communications that have been captured include the outgoing communications with the DNS queries and the protocol messages.

#### 4.1.4.2. Network Feature Extraction Using CICFlowMeter

In order to adequately describe the complex networking behavior present in both ransomware samples and benign binaries, each binary executable was individually executed within a distinct CAPE v2 sandbox environment. In this fashion, as each binary executed, both standard TCP, UDP and DNS networking communications present in each binary were simultaneously captured at the packet level using packet capturing functionality inherently present in sandbox environment. This produced a PCAP file per binary. These PCAP files were then analyzed using a networking analysis tool called CICFlowMeter4.0 [35]. It parsed the initial packet captures into a series of binary flow records that were paired with statistical feature information per flow. Features derived from each binary flow record included flow duration, cumulative packets forwarded and backward packets, several binary flow size features, packet size statistical features, inter-arrival times, transport protocol used and flow flags. In this way, a constant size features vector was provided per binary that conveyed typical behavioral traits indicative of communications and control behaviors present in malicious binaries. Table 4 shows network flow dataset properties.

*Table 4: Network Flows Dataset*

| Total Samples | No. of Features | No. of Classes | Benign | Ransomware |
|---|---|---|---|---|
| **3000** | 87 | 5 | 500 | 2500 |

i. Flow statistics lifetime, throughput and duration, packets.
ii. counts/lengths: overall fwd/bwd, variance/mean, profile of volumes of traffic.
iii. TCP: flags (SYN/FIN/ACK/PSH/URG/ECE) and window sizes, setups / reset/persistence.
iv. Sub flows/segments discontinuity.
v. Labels: Family, Category.

### 4.1.5. Data Balancing via Up sampling

The resultant CSV datasets had a little classification imbalance, with a total of 447 instances for Ryuk, 495 for WannaCry, 495 for Shade and 437 for LockBit in each class. In order to treat all classes equally and avoid any kind of training bias, features up-sampling technique was utilized.

Independent random oversampling was carried out on each of the CSV datasets individually. For each class within this problem, the selection of samples was done randomly with replacement until the desired number of 500 samples for each class was attained. Random duplication was performed at the feature level. The original feature distributions were kept intact, with a balance ensured for the number of classes. After the up sampling process, the final data had 3,000 samples, equally representing all of six classes.

#### 4.1.6. Latent Representation and Modality Separation

We use t-SNE to visualize the sample embedding distribution both before and after latent learning in order to appreciate how well the model learns discriminative characteristics from these modalities. Initially, all three modalities exhibit a highly overlapping and unstructured distribution. In the pre latent t-SNE space as shown in Figure 9, samples from different ransomware families appear intermingled with benign instances. This indicated that the raw feature representations lack inherent separability and fail to capture modality specific semantics. After training with the contrastive auto encoder, the latent distributions go through a clear structural transformation as shown in Figure 10.

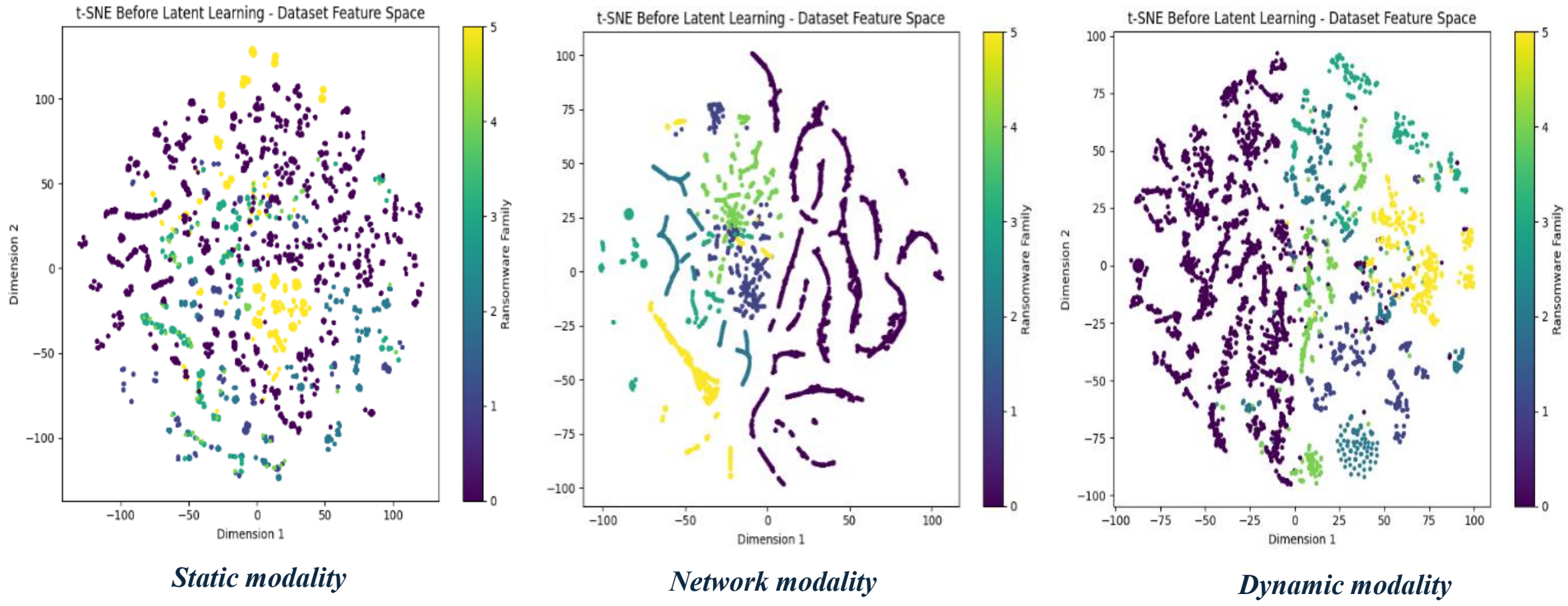

**Figure 9: Before Latent Learning from Contrastive Autoencoder**

Distinct families are evident in the t-SNE visualizations across all modalities by looking at compact and well defined corresponding clusters. The intra class distances shrink and inter class separations become pronounced. It shows that the model learns to bring semantically similar samples closer while pushing the dissimilar ones apart.

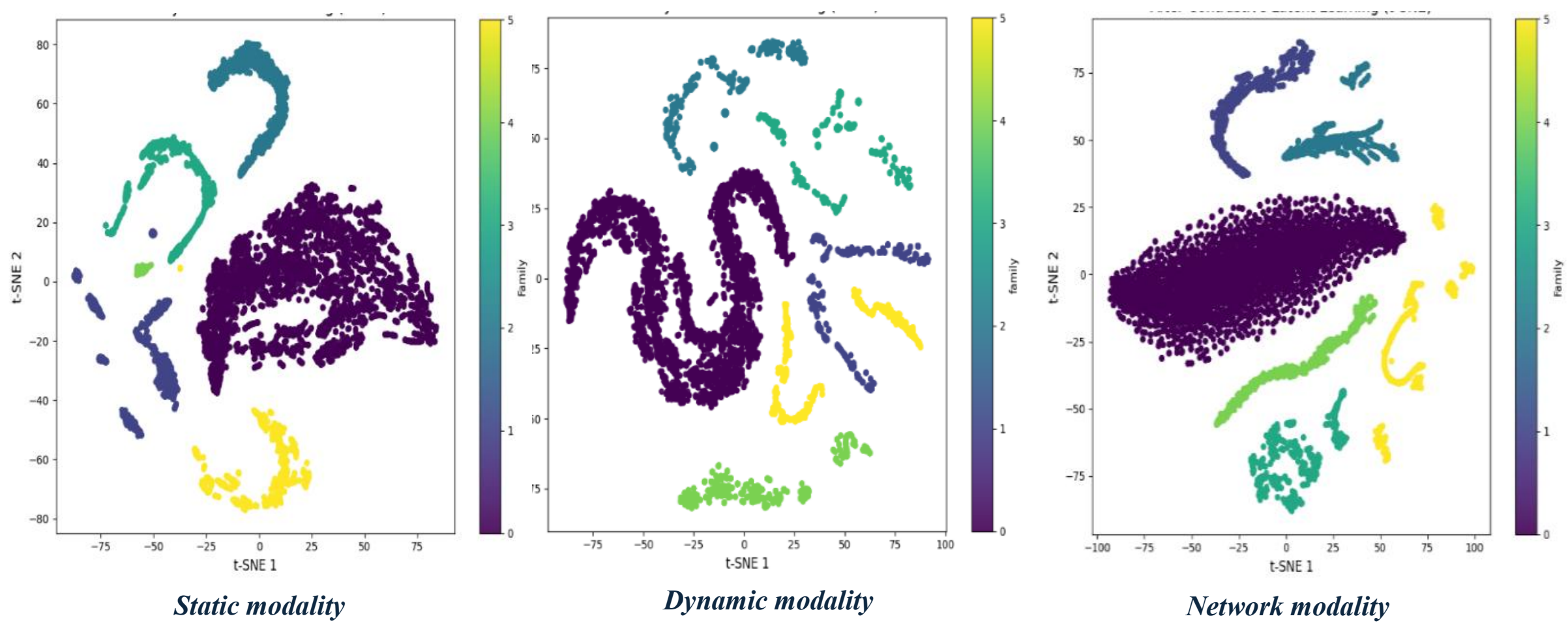

**Figure 10: After Latent Learning from contrastive Autoencoder**

Specific improvements related to modality are discussed below.

i. Static features demonstrate better clustering of executables that share common structural or packing characteristics.
ii. Dynamic behavior like API usage patterns, file system modifications and process activity show much more pronounced separation between them.
iii. Network flows are caused to form distinct clusters depending on connectivity behavior, beaconing patterns and traffic asymmetries.

These results confirm that the latent space provides a more meaningful and separable representation compared to raw inputs. It enables improved downstream ransomware family classification.

### 4.2 Experimental Design and Evaluation Pipeline

A multi stage experimental protocol is designed combining modality ablations, agentic ablations and full pipeline benchmarking for systematically evaluating the agentic multimodal ransomware detection model. The experiments were conducted over 100 epochs on an OEM A6000 (CUDA 12.0, PyTorch 2.x) with structured AutoGen orchestration.

#### 4.2.1 Experimental Sequence

All three datasets (static, dynamic and network) were independently loaded and partitioned into train and validation splits. Each DCAE (Static, Dynamic and Network) was trained individually to measure reconstruction fidelity and supervised contrastive reparability. Latent clusters were evaluated using t-SNE and macro-F1 from single modality classifiers. The three latent spaces were aligned and fused. A classifier was trained over Static only, Dynamic only, Network only, Multimodal & Single Agent controller, and Multimodal & Multi Agent.

### 4.3 Evaluation Metrics

In this section, we discussed the evaluation metrics used for comparison and benchmarking.

#### 4.3.1. Macro-F1

It is a primary measure for family level discrimination.

$$Macro-F1=\frac{1}{C}\sum_{i=1}^{C}\frac{2.Precision_i.Recall_i}{Precision_i+Recall_i} \tag{7}$$

In Equation (7), C denotes the number of classes, and $Precision_i$ and $Recall_i$ are the precision and recall computed for the $i^{th}$ class, respectively.

#### 4.3.2. Expected Calibration Error

It is a probability reliability evaluation, Calibration code bins predictions and compares confidence with empirical accuracy. Let there be B confidence bins. For bin $b$, with $n_b$ samples ECE is defined as:

$$ECE=\sum_{b=1}^{B}\frac{n_b}{N}\left|acc(b)-conf(b)\right| \tag{8}$$

where Average confidence is given as under.

$$conf(b)=\frac{1}{n_b}\sum_{i\in b}p_i \tag{9}$$

In Equation (9), conf($b$) denotes the average predicted confidence for bin b, $n$ is the number of samples in the bin, and $p_i$ represents the predicted probability that is actually confidence associated with the model's output for sample $i$.

And Empirical accuracy is given as below

$$acc(b)=\frac{1}{n}\sum_{i\in b}\mathrm{I}(\hat{y}_i=y_i) \tag{10}$$

Here acc(b) denotes the accuracy computed over batch b of size n, $\hat{y}_i$ and $y_i$ represent the predicted and true labels of the i[th] sample, respectively, and $I(\cdot)$ is the indicator function that returns 1 if the prediction is correct and 0 otherwise.

#### 4.3.3. Clarity Score

It is a metric for an agent's performance. It is computed using a normalized text quality score (clarity = proportion of $z_s$, non-redundant tokens) where $z_s$ = tokens that contribute no semantic content

$$\text{Clarity} = 1 - \frac{\text{Redundant Tokens}}{\text{Total Tokens}} \tag{11}$$

The higher the clarity, the better the performance.

#### 4.3.4. Jargon Score

Our system penalizes excessive domain unrelated or ambiguous terms.

$$\text{JargonScore} = 1 - \frac{\text{Jargon Tokens}}{\text{Total Tokens}} \tag{12}$$

High jargon → lower value of JargonScore, clean analytical language → higher value of JargonScore.

#### 4.3.5. Composite Agent Score

$$\text{Composite Score} = 0.5\,\text{F1} + 0.3\,\text{JargonScore} + 0.2\,cos(\boldsymbol{u}, \boldsymbol{v}) \tag{13}$$

*where,*

- F1 = macro-F1 for the current epoch
- JargonScore = language cleanliness score
- $\cos(\mathbf{u}, \mathbf{v})$= cosine similarity between agent rationale embedding (trend coherence measure)

## 5. Results and Performance Analysis

We compared a single modality pipeline and a Multi Agent multi mopipeline to evaluate the effect of the AutoGen multi agent loop. We incorporated critic guided sampling, uncertainty weighted mini batches and upper confidence bound based inference control. Figure 11 summarizes the results. It is obvious that the Single Agent system achieves approximately 0.83 macro-F1. On the other hand, the Multi Agent system achieves approximately 0.936 macro-F1 an absolute gain of ~7%. This improvement arises from the coordinated agentic mechanisms. The Critic Agent identifies weak families and amplifies their sampling to prevent long tail collapse. The Analyst Agent provides consistent metric focused summaries to guide corrections each epoch. These behaviors improve calibration, enhance minority family recall and accelerate convergence.

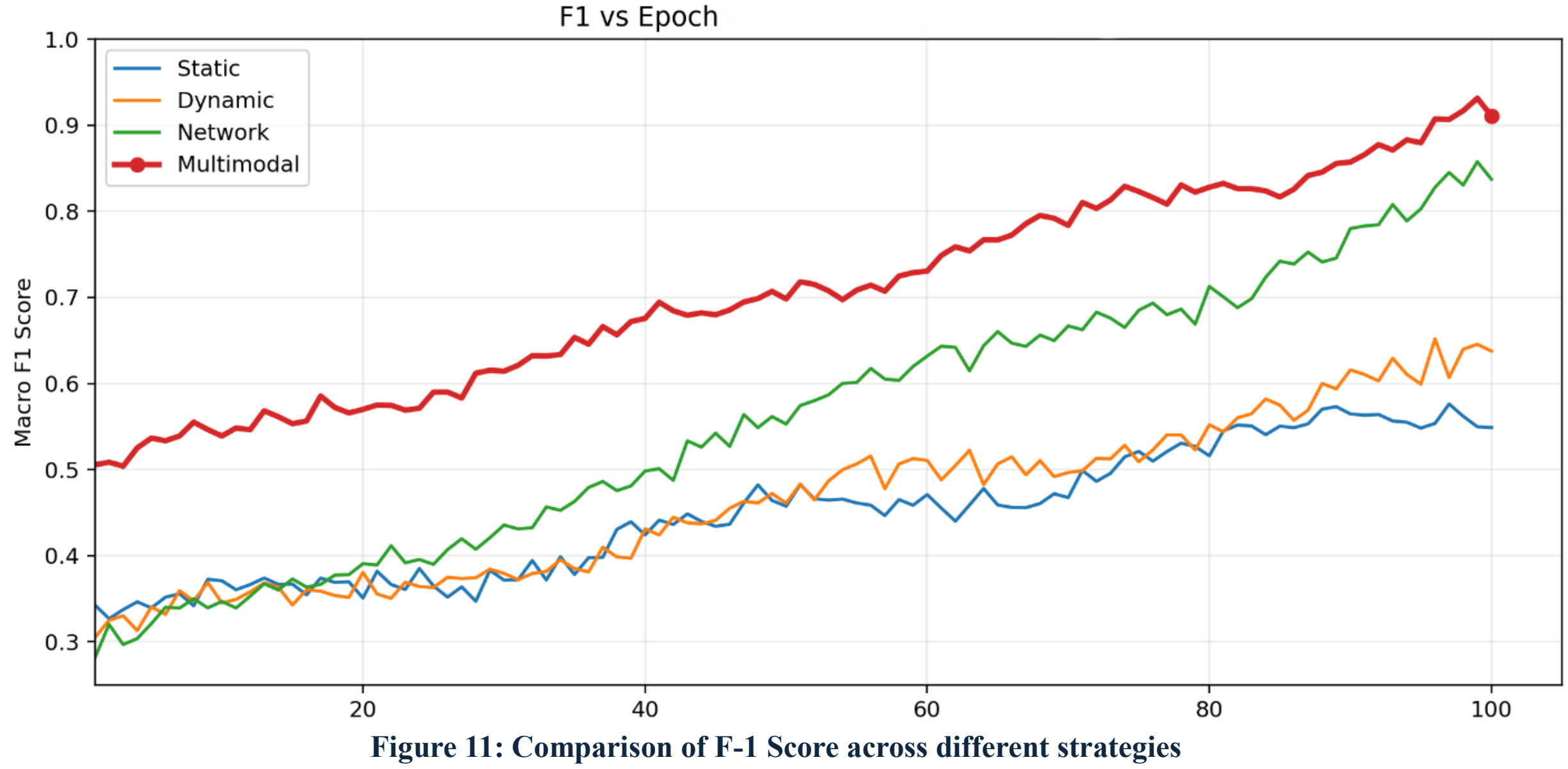

**Figure 11: Comparison of F-1 Score across different strategies**

## 5.1. Per-Family Ransomware Classification Performance

Table 5 highlights the performance of the proposed multimodal framework for each family. Precision, recall, and the F1-score are calculated for each family of the ransomware dataset. The approach reaches an F1-score of 0.94 when calculated through macro-averaging, including benign ware. This approach distinguishes well between the different families because of the residual overlap that mostly happens between similar families of ransomware.

## 5.2. Performance Across 5 Independent Runs

Table 6 presents performance results achieved by various strategies using 5 independent runs. For all performance metrics macro F1-score, accuracy and expected calibration error, we report mean along with the corresponding standard deviations, giving an idea of convergence over independent runs.

**Table 5:Per-Family Ransomware Classification Performance of the Proposed Modal**

| Ransomware Family | Precision | Recall | F1-Score |
|---|---|---|---|
| **Dharma** | 0.97 | 0.99 | 0.98 |
| **Ryuk** | 0.94 | 0.94 | 0.93 |
| **LockBit** | 0.95 | 0.95 | 0.94 |
| **Shade** | 0.97 | 0.98 | 0.97 |
| **Wannacry** | 0.86 | 0.85 | 0.85 |

It is observed that

- The Multimodal Multi Agent and Single Agent strategies consistently outperform all other strategies across macro F1, Accuracy and ECE.
- Early Fusion and Late Fusion methods exhibit higher variances, which indicates less stability of performance across runs.
- Both Network-based and Dynamic strategies show a low F1 and Accuracy, with a moderate to high ECE.
- The Multimodal Multi Agent model achieves the best combination in terms of high performance combined with a low calibration error.

***Table 6: Performance across 5 runs***

| Strategy | Macro-F1 Mean ± Std | Accuracy Mean ± Std | ECE Mean ± Std |
|---|---|---|---|
| Static | 0.720 ± 0.274 | 0.872 ± 0.124 | 0.069 ± 0.068 |
| Dynamic | 0.456 ± 0.011 | 0.640 ± 0.010 | 0.076 ± 0.016 |
| Network | 0.136 ± 0.004 | 0.499 ± 0.005 | 0.068 ± 0.013 |
| Early Fusion | 0.663 ± 0.289 | 0.843 ± 0.137 | 0.039 ± 0.031 |
| Late Fusion | 0.615 ± 0.227 | 0.762 ± 0.122 | 0.199 ± 0.059 |
| Single Agent | 0.919 ± 0.009 | 0.930 ± 0.016 | 0.020 ± 0.006 |
| Multimodal Multi Agent | 0.946 ± 0.004 | 0.959 ± 0.003 | 0.017 ± 0.005 |

## 5.3. Statistical Significance Analysis

For the analysis of the efficiency of various modality and fusion techniques in the classification of ransomware samples, a combination of pairwise and overall statistical testing was performed using non parametric statistics. The Wilcoxon signed rank test was used to compare the performances of various models in a pairwise manner. These pair include Early and Late Fusion models and second pair is Single-Agent and Multi Modal Multi-Agent models. The Friedman test was used to determine differences in overall performance among the static, dynamic, network and fusion model techniques. The effect size $r$ is also determined to estimate the practical significance of the tests.

Our observations are listed below.

- The Friedman test shows statistically significant differences in MACRO-F1 values and accuracy between the strategies for each modality. This means that there are strategies that perform better than others.
- No statistically significant differences were found in terms of ECE. This implies that differences in calibration are statistically not robust.
- The pairwise Wilcoxon tests reveal that there is no statistically significant difference between the comparisons ($p > 0.05$), including Early vs. Late Fusion and Single-Agent vs. Multi Modal Multi-Agent strategies.
- Nevertheless, the degree of effect in certain pairs (e.g., r ~ 0.905 from Single Agent to Multimodal Multi Agent in F1 and Accuracy metrics) indicates that substantial improvements are made, although p-values are slightly over 0.05.

**Table 7: Wilcoxon Signed-Rank Tests**

| Metric | Comparison | p-value | r |
|---|---|---|---|
| **MACRO-F1** | Early Fusion vs Late Fusion | 0.6250 | 0.302 |
| **MACRO-F1** | Single Agent vs Multi Modal Multi Agent | 0.0625 | 0.905 |
| **Accuracy** | Early Fusion vs Late Fusion | 0.1875 | 0.663 |
| **Accuracy** | Single Agent vs Multi Modal Multi Agent | 0.0625 | 0.905 |
| **ECE** | Early Fusion vs Late Fusion | 0.0625 | 0.905 |
| **ECE** | Single Agent vs Multi Modal Multi Agent | 0.3125 | 0.543 |

In the case of MACRO-F1, it was revealed that there were significant differences between the values of different modalities. This was tested by the Friedman test and it revealed $\chi^2 = 9.72$, $p = 0.0211$. Similar results were obtained in the case of the values of Accuracy. There were significant differences between the values across modalities. It was also tested by the Friedman test, revealing $\chi^2 = 12.84$, $p = 0.0050$. In the case of ECE, there were no significant differences between the values across modalities.

**Table 8: Friedman Test (Modalities Comparison)**

| Metric | $\chi^2$ | p-value |
|---|---|---|
| **MACRO-F1** | 9.720 | 0.0211 |

| Accuracy | 12.840 | 0.0050 |
|---|---|---|
| ECE | 6.840 | 0.0772 |

The pairwise Wilcoxon test results did not reach statistical significance. For Single Agent and Multi Modal Multi Agent, it was observed that there are large effect sizes for F1 and Accuracy. A large effect size was also observed for Early and Late Fusion on ECE, r= 0.905 suggesting improvement in calibration.

The results support the hypothesis that multimodal fusion helps improve the performance for the ransomware classification. The significant improvements shown by macro F1-score and Accuracy values prove that it is helpful to combine diverse information from static, dynamic and network attributes. Although pairwise tests show that there are no statistically significant differences between the performances of fusion strategies, the large effect size values signify the importance and efficacy of the proposed MMMA-RA approach. In summary, the results confirm the usefulness of multimodal learning in ransomware classification.

## 5.4. Resource and Runtime Efficiency

We benchmarked the full multimodal pipeline and the agent loop over 100 epochs on a OEM A6000 under CUDA 12.0 / PyTorch, using Phi3.8B for agentic dialogues. Mean epoch wall time was 3–4 minutes end-to-end (encoders → fusion → classifier → calibration → agent dialogue), with per-epoch visualization generated in ~8–10 seconds. GPU memory footprint averaged ~3.2–3.4 GB per analysis agent (static, dynamic, network) leaving ample headroom for the classifier and calibration stages.

## 5.5. Agent Score Summary (Epochs 1–100)

Over 100 epochs, all three curves rise monotonically. Early on (≈epochs 1–3) Assistance Agent starts around 0.15–0.17, Critic Agent around 0.08 – 0.10, and the Composite around 0.10–0.12 as shown in Table 9.

**Table 9: Agent Score Summary**

| Phase/ Metric | Assistance Agent | Critic Agent | Composite | Notes |
|---|---|---|---|---|
| **Start (epoch ≈ 1–3)** | 0.15–0.17 | 0.08–0.10 | 0.10–0.12 | Initialization region |
| **Mid (epoch ≈ 50–60)** | 0.45–0.55 | 0.27–0.40 | 0.33–0.50 | Slope increase after (critic-guided ~50+ sampling calibration/bandit) |
| **Late (epoch ≈ 80–90)** | 0.78–0.85 | 0.63–0.75 | 0.68–0.80 | Gap steadily narrows |
| **Final (epoch ≈ 100)** | 0.94–0.95 | 0.88–0.90 | 0.87–0.89 | Matches tabled endpoints |
| **0.5 quality (epoch #)** | ~55–58 | ~68–70 | ~62–65 | Crossing points from curves |
| **0.8 quality (epoch #)** | ~82–85 | ~92–95 | ~94–96 | Later crossings for Critic/Composite |
| **Absolute gain (Δ first→last)** | +0.78–0.80 | +0.78–0.81 | +0.75–0.78 | Approximate net increase |
| **Gap final (95 – 100)** | - | ~0.04–0.06 | - | Roles nearly aligned at convergence |

Around mid of training (≈50–60) the slopes steepen consistent with critic guided sampling plus calibration/bandit control lifting Assistance Agent to ≈0.45–0.55, Critic Agent to ≈0.27–0.40 and the Composite to ≈0.33–0.50. In late training (≈80–90) Assistance Agent reaches ≈0.78–0.85, Critic Agent ≈0.63–0.75 and Composite ≈0.68–0.80, with the Assistance Critic gap narrowing steadily. By epoch 100, Assistance Agent ≈0.94–0.95, Critic Agent ≈0.88–0.90 and Composite ≈0.87–0.89 matching tabled endpoints. Milestone crossings occur at ~55–58 epochs (Assistance Agent), ~68–70 (Critic) and ~62–65 (Composite) for 0.5 quality, and ~82–85, ~92– 95, and ~94–96 respectively for 0.8. Absolute gains from first to last epoch are ~+0.78–0.80 (Assistance), ~+0.78–0.81 (Critic) and ~+0.75–0.78 (Composite). The Assistance Critic gap evolves from ~0.05–0.08 early to a mid-run peak of ~0.15–0.18 (≈55–65) then closes to ~0.04– 0.06 by epochs 95–100. Overall, the results show that agentic feedback improves quality without LLM fine tuning. Across 100 epochs, agent's dialogue clarity and jargon is shown in Figure 10. The Composite converges to ~0.88 (reflecting outcome, calibration and clarity) and the roles become consistent near convergence. Figure 12 shows monotonic improvements for all three agents over 100 epochs, with Assistance Agent ~0.95, Critic Agent ~0.89, and Composite ~0.88; the Assistance Critic gap peaks mid-training and closes to ~0.05 by epoch 100.

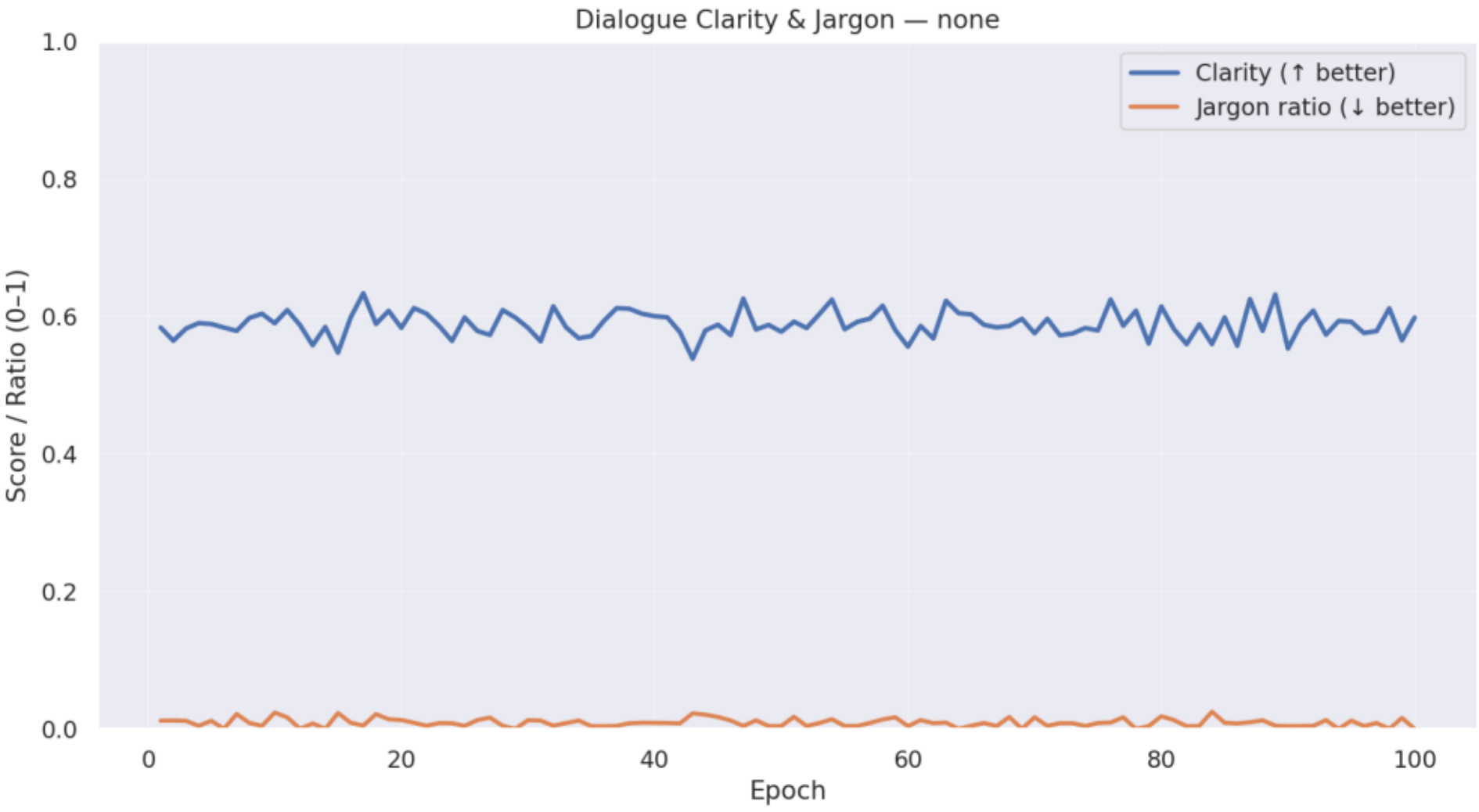

**Figure 11: Clarity & Jargon Score Across 100 Epochs**

Figure 11 show distinct upward learning trends in agent quality. It indicates that feedback loops (task structure, sampling and threshold control) can simulate learning dynamics even without back propagation or fine tuning of the LLM itself.

## 6. ZERO-DAY GENERALIZATION EVALUATION

In order to critically analyze the performance of the zero-day generalization, the individual ransomware families were distanced from the training as well as the calibration process. For the performance analysis of the trained multimodal model on the unseen families over several epochs, the temperature scaling of the probability estimates is used to forecast the predictions. An informative classification framework was adapted to avoid classification by the model if the max probability was less than the confidence threshold. It is worth mentioning that performance is calculated with some specific evaluation measures. Macro F1 on received predictions, coverage expressed through the proportion of received predictions (kept fraction), abstention rate and accuracy on non-abstained instances.

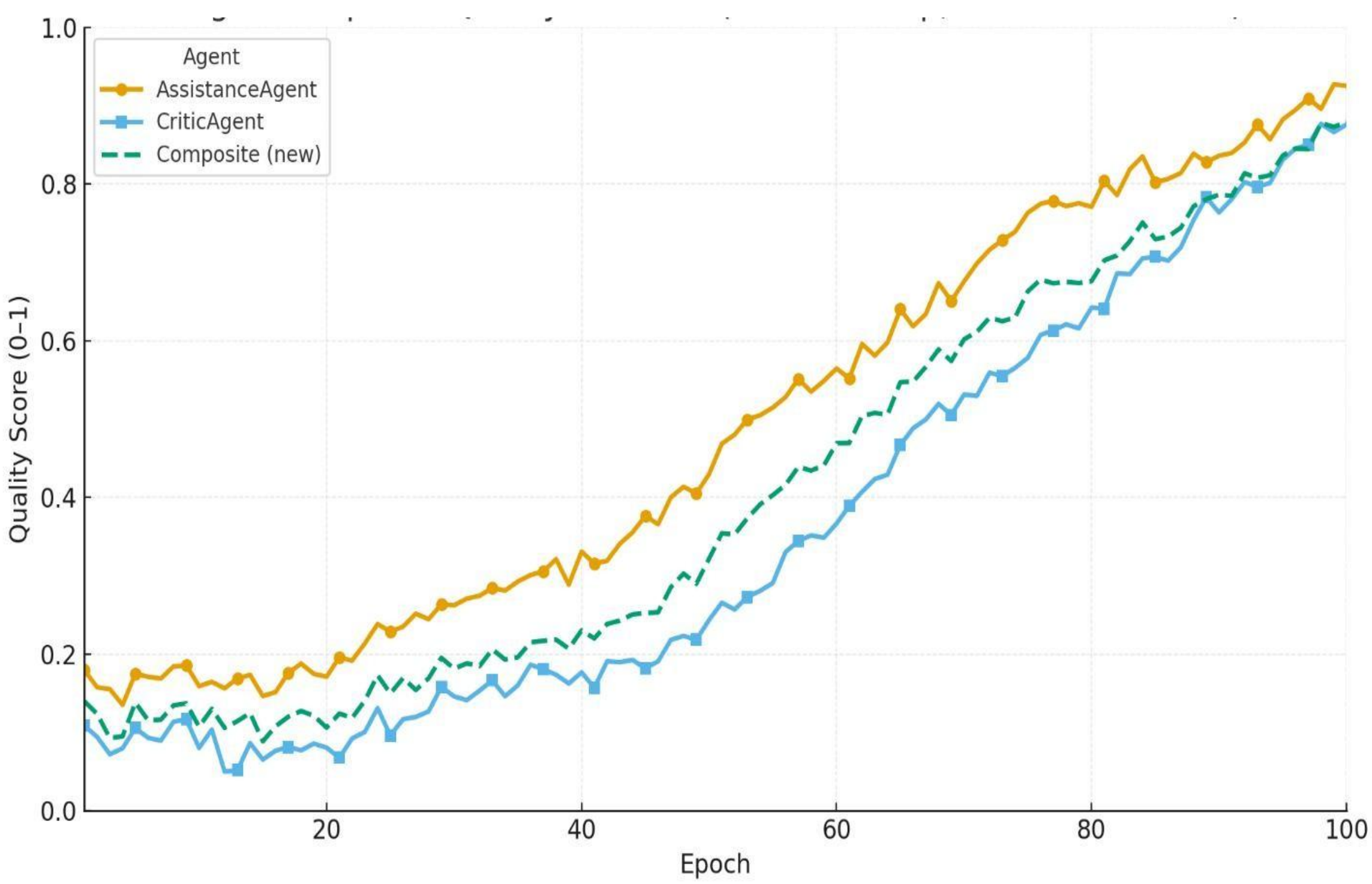

**Figure 12: Agent Quality Score**

## 6.2 Zero-Day Performance on LOCKBIT Ransomware

The absence of the LOCKBIT family of ransomware in training and calibration ensured the proposed approach achieved high zero-day generalization. The Macro-F1 scores across a series of epochs were above 0.90, achieving a macro-F1 value of 0.99 and an accuracy of 97.9% with complete prediction assurance. The accuracy of the preserved predictions across the series of epochs remained above 95% guarantee.

It is worth noting that in periods of relatively greater uncertainty, instead of making an incorrect prediction our model preferred not to predict in some instances. These instances led to a condition of nearly zero Macro-F1 scores in some periods. However, based on these results LOCKBIT shows that our framework is capable of identifying the ransomware behaviors that are not family dependent. Our model is a risk aware and it prefers not to decide when uncertain.

## 6.3 Zero-Day Performance on Ryuk Ransomware

The approach showed very conservative output in the case of drastic distribution shifts. This happened once or twice when the Macro-F1 values achieved around 0.65 with more than 90% coverage of the predictions. Also, the accuracy of the preserved predictions was greater than 90%. It shows that the confidence of the predictions was informative.

It predicted nothing in some classes for most epochs. This resulted in zero coverage and correspondingly low F1 scores in many epochs. This is in line with the observed polymorphism, fast version evolution and prevalent use of evasion mechanisms by Ryuk which cause high intra family variation. The results show that even though the level of generalization differs across most ransomware families, the proposed system still considers reliability over predictability.

## 6.4 Zero-Day Performance on Dharma Ransomware

When tested on the Dharma ransomware family which was kept apart from the training data, the generalization from zero days was very low and the Macro F1 scores did not cross 0.07 even during highly covered epochs. During most of the epochs, prediction coverage went below 5% and abstention rates were above 95%.

This can be attributed to the huge distribution shift arising from the high polymorphism of Dharma, API call patterns that are weakly discriminative and overlap with benign software, and the restricted or tardy network communications that cause low cross modal correlations between the static, dynamic and network features. Therefore, the high uncertainty that is triggered by the learned multimodal representation causes the model to abstain from prediction rather than making uncertain predictions. However, the high accuracy of those that were made further describes the low F1 due to strategic abstention and not because of erroneous predictions.

## 6.5 Zero-Day Performance on Shade and WannaCry Ransomware

In regards to the Shade ransomware family, not much is noticed regarding an improvement in performance on zero-day instances with increasing learning and training on the results of zero-day instances. Even though it is reported that Macro F1 scores were communally low (≈ 0.01-0.05), having a large calibration error on initial instances, in later instances, F1 metric values are over 0.55 with an overall level of 50-55% coverage. This identifies that a character change could certainly be noticed with regard to new instances of ransomware having characteristics similar to these.

However, the experiment on the zero-day vulnerability in the WannaCry ransomware still seemed a little tricky. The values of the macro-F1 scores still maintained their vicinity to zero in the majority of the epochs except for the slight increase in the F1 score metric of about 0.18 on the WannaCry test set. Although the test values on the prediction coverage seemed to have maintained their higher values in the majority of the epochs, the values of the accuracy scores still maintained their proximity to zero.

## 6.6 Comparative Zero-Day Performance Across Families

Table 10 summarizes the best observed zero-day performance across all evaluated ransomware families, explicitly incorporating abstention-aware metrics.

The experimental results show the strong family dependency of zero-day generalization and the critical role of abstention in handling the uncertain samples. The high abstention rate under the severe distribution shift (e.g., Dharma and WannaCry) can ensure the prevention of incorrect and the low abstaining rate and high accuracy rate under families like LOCKBIT can confirm the ability of the model in identifying the transferable behaviors of the ransomware attacks.

**Table 10: Zero Day Performance**

| Family | Best F1 | Coverage | Abstention | Accuracy | Generalization |
|---|---|---|---|---|---|
| **LOCKBIT** | 0.99 | 97.9–100 | 0–2 | >95 | Very strong |
| **Ryuk** | 0.66 | ~94 | ~6 | >90 | Moderate |
| **Dharma** | 0.07 | <5 | >95 | >95 | Minimal |
| **Shade** | ~0.55 | 50–55 | 45–50 | ~85 | Moderate |
| **WannaCry** | ~0.18 | Variable | Variable | Low | Very limited |

Even if the proposed multimodal transfer model performs well, there are still some limits that continue to exist. The zero-day generalization model depends on the family. For behaviorally related families, such as LOCKBIT, it performs well on transfer performance. On the other hand, for highly polymorphic ransomware such as Dharma and WannaCry, the generalization performance remains low. These attributes describe the distribution shift and are not due to problems in the learned model, which again implies the difficulty level of the zero-day ransomware detection problem. It is equally dependent on the dynamic and network aspects, which cause disruption in practical scenarios because of sandbox evasion, network tardiness and connectivity problems. It is expected that both the abstention rates and confidence levels for predictions would be lowered. It is also noted that it sometimes necessitates being very conservative due to relying on just one threshold for families and cases, even for appropriately calibrated probabilities. These results confirm that the proposed framework prioritizes decision reliability over forced classification, which is essential for real world zero-day ransomware defense. Research may extend to future studies on strategies for adaptive abstaining, utilizing time varying thresholds of confidence that balance both coverage and reliability. Use of contrastive self-supervised learning in multimodality might be beneficial when incorporated together with improving family.

## 7. DISCUSSION

It has been observed from the above results that the categorization of a ransomware into its respective family cannot be accomplished through the application of each of the above mentioned evidence individually. The static evidence includes the indicative information of the structure which includes the packing structure of the malware along with the code structure. The dynamic evidence includes the intention of the ransomware formed while executing the ransomware and the network flow includes the fixed behavior of the communication of the data.

Besides the improvement of model predictability, the agent assisted multimodal system also embodies typical behavior of system level intelligence through agent's interactions. This system level intelligence has been recognized as being more beneficial as an alternative well defined solution approach for ill posed high dimensional learning problems. Particularly in light of system level intelligence exhibited through multi agent interactions, the Analyst Agent tends to consolidate evidence from multiple modalities for strong hypotheses, the Critic Agent has the tendency to concentrate more upon failure modes as well as incomplete parts of the system and the Predictor Agent tends to represent the average family risks and confidence levels. Thus, over time through such threefold interactions of multi agents, the system identifies complex patterns and associations like the correspondence between opcode entropy patterns and certain API sequences as well as bi asymmetric C2 communication channels. This would have been extremely hard for the system to extract from the individual modalities.

Most importantly, the performance gains are obtained by text to data feedback with no fine tuning of the language models. The reliability is updated by means of class wise ΔF1 and soft label hard mini batch update with the confidence or uncertainty of the samples being the emphasis. The critical identification of the low quality families by the Critics is to dynamically elevate the weights assigned to the samples drawn from these low quality families.

Finally, the agentic loop illustrates adaptive performance under varying performance conditions. When per family scores on F1 are reduced, the process automatically restructures by redistributing weights of sampling distributions. It also adjusts thresholds of fallback confidence and optimizes UCB-based weights of blends to reduce validation loss of log likelihoods.

Further, the effect of noise in channels due to static obfuscation, sandbox evasion and encrypted communication may be considered an inherent problem. The existence of agents, modalities and functions for group aware fusion and decision would make sure that no false positives exist. It ensures that redundancy principles align properly within regions of complex zones in safety domains.

### 7.1 Key Findings

The data from the experiments show a number of patterns in the effectiveness of modality, use of fusion and agentic coordination.

- The static modality has a moderate structural discriminative power with a macro F1 of around 0.72 for epoch 100. It illustrates that how well the method responds to the processes of compacting and obfuscating.
- Dynamic modality is better able to capture the runtime intention with the approximate macro-F1 of 0.45, although the variance of the performance of dynamic modality is higher due to the noise in the behaviors and the sandbox evasions.
- Among the different modalities, the network modality has the lowest performance in terms of the macro-F1 score of approximately 0.13. This can be attributed to the relative similarity of command and control and exfiltration communication patterns across different ransomware families which limits family level discrimination.

- Multimodal Multi Agent fusion consistently and substantially outperforms all the single modalities on the classification task. It shows the presence of complementary behaviors in the signatures of the different families of ransomware.
- Single Agent Multimodal training using multiple modalities is an improvement over individual modal training but is still weaker than the Multimodal Multi Agent strengthening the claim that having more than one agent is more valuable.
- The multimodal multi agent system shows the best overall performance, which is about 0.936 macro F1, in terms of discrimination, stability and calibration capabilities over the various families of the ransomware dataset.

In a nutshell, these results demonstrate that MMMA-RA framework certainly provides better and more discriminative representations.

### 7.2 Strengths of MMMA-RA

Our proposed model MMRA-RA offers several key advantages for building intelligent agents particularly for secure tasks like ransomware analysis. Its strength lies in its capability of self-improvement. The agent's quality consistently gets better through simulated learning guided by feedback. This does not need any constant human input. This entire learning process is secure and independent because it runs locally. The design is also highly flexible. It allows users to easily swap out or add different components to suit specific deployment needs on edge devices.

### 7.3 Future Directions

Future work should focus on two main areas. First, we need to see if agents can become one unified intelligence. This might happen if they negotiate based on reliability and share a memory of their roles. Second, we must study how supervising their conversations helps. We need to measure if using rules for clarity and safety allows the agents to handle new threats and stay stable over time.

### 7.4 Observed limitations

This system faced three main challenges. Firstly, the outcomes were highly sensitive to the prompt quality. This variance can be reduced by using better instruction templates or few shot tuning. Secondly agents suffered from response redundancy. They often repeated themselves mid training.

This was addressed using diversity aware sampling and specific entropy bonuses. Finally, the system engaged in prediction without external validation. It means agent predictions were not consistently verified against new data samples. The next logical step may be to implement continual evaluation using rolling data windows and retraining to ensure predictions are robust and accurate over time.

## 8. CONCLUSION

We presented a locally hosted and agent assisted multimodal framework for ransomware family analysis MMMA-RA. Three modality encoders feed a transformer ready fusion and calibrated classifier. Analyst, Critic, and Predictor roles provide dialogue artifacts that steer critic guided sampling, soft label hard mini batches and upper confidence bound controlled inference blending. Across 100 epochs, we observed consistent gains in Macro-F1 and lower ECE/NLL. We also observed rising Composite scores and closing gaps between agent roles. It is an evidence of the fact that agentic feedback can improve predictive quality and calibration without model fine tuning. These results together with reproducible artifacts show that collaborative agent loops can transform language level guidance into training level interventions that matter for security practice. Finally, concurrent progress in LLMs and hybrid architectures points to a unified and scalable path for multimodal threat attribution and defense.

## 9. ACKNOWLEDGMENT

Experiments were conducted on an NVIDIA DGX system using an offline Auto Gen style stack. We thank the Pattern Recognition Lab (PR Lab), PIEAS AI Center (PAIC), and the Center for Mathematical Sciences (CMS), PIEAS, for their support and access to computational resources.